\documentclass[prb,reprint,showpacs,amsmath,amssymb,floatfix,footinbib]{revtex4-1}
\usepackage{graphicx}
\usepackage{color}
\usepackage{pifont}
\usepackage{amstext}
\usepackage{amsmath}
\usepackage{bm}
\usepackage{graphics}
\usepackage{tabularx}
\usepackage{amssymb}
\usepackage{wasysym}
\usepackage{etoolbox}
\usepackage{colortbl}
\usepackage{xspace}
\usepackage{subfigure}

\setcitestyle{square,numbers}
\newtoggle{clearpage}
\togglefalse{clearpage}

\newcommand{\ie}[0]{i.e.\@\xspace}
\newcommand{\eg}[0]{e.g.\@\xspace}
\definecolor{darkgreen}{rgb}{0.0,0.5,0.0}

\begin{document}
%
\title{Kane-Mele-Hubbard model on the $\pi$-flux honeycomb lattice}
\author{Martin \surname{Bercx}}
\email{martin.bercx@physik.uni-wuerzburg.de}
\author{Martin \surname{Hohenadler}}
\author{Fakher~F. \surname{Assaad}}
\affiliation{ Institut f\"ur Theoretische Physik und Astrophysik,
Universit\"at W\"urzburg, Am Hubland, 97074 W\"urzburg, Germany }
\begin{abstract}
  We consider the Kane-Mele-Hubbard model with a magnetic $\pi$ flux
  threading each honeycomb plaquette. The resulting model has remarkably rich
  physical properties. In each spin sector, the noninteracting band structure
  is characterized by a total Chern number $C=\pm 2$. Fine-tuning of the
  intrinsic spin-orbit coupling $\lambda$ leads to a quadratic band crossing
  point associated with a topological phase transition. At this point,
  quantum Monte Carlo simulations reveal a magnetically ordered phase which
  extends to weak coupling. Although the spinful model has two Kramers
  doublets at each edge and is explicitly shown to be a $Z_{2}$ trivial
  insulator, the helical edge states are protected at the single-particle
  level by translation symmetry. Drawing on the bosonized low-energy
  Hamiltonian, we predict a correlation-induced gap as a result of umklapp
  scattering for half-filled bands. For strong interactions, this prediction
  is confirmed by quantum Monte Carlo simulations.
\end{abstract}
 \pacs{71.10.-w, 03.65.Vf, 73.43.-f,71.27.+a}
\maketitle
%
\section{Introduction}\label{sec_intro}
The classification of insulating states of matter has been refined in terms of protecting symmetries through 
the discovery of topological insulators \cite{KaneMele05a,KaneMele05b,Bernevig06,König07}.
For example, as long as time-reversal symmetry is not
broken, topological insulators cannot be adiabatically connected to
nontopological band insulators without closing the charge gap \cite{Schnyder08}, 
and the helical edge states are protected against
perturbations \cite{KaneMele05a,KaneMele05b,Wu06,Xu06}.

Recently, a further refinement was achieved by the theoretical prediction
\cite{Fu11, Hsieh12,Slager13} and experimental realization
\cite{Tanaka12,Dziawa12,Xu12} of topological crystalline insulators
(TCIs). In this case, in addition to time-reversal symmetry, the
two-dimensional surface has crystal symmetries which protect the topological
state against perturbations. Because crystal (point group) symmetries are not
defined in one dimension, this definition of TCIs requires a three-dimensional
bulk and a two-dimensional surface.

Here, we introduce a two-dimensional counterpart to the TCI. In addition to
time-reversal symmetry, the model we consider preserves translation symmetry
at the one-dimensional edge.
This leads to  protection at the single-particle
level despite a trivial bulk $Z_2$ invariant. Our model is based on the
Kane-Mele (KM) model \cite{KaneMele05a} on the honeycomb lattice, which has a
quantum spin Hall ground state at half filling.
 By threading each honeycomb
plaquette with a magnetic flux of size $\pm\pi$, we obtain the {\it $\pi$
  Kane-Mele ($\pi$KM) model}. The idea of inserting $\pi$ fluxes has
previously been considered for the case of an intensive number of fluxes
\cite{Ran08,Qi08,Juricic12,Assaad13}, and a superlattice of well separated
fluxes \cite{Wu13}. Isolated magnetic $\pi$ fluxes locally bind zero-energy
modes and lead to spin-charge separation in topological insulators
\cite{Ran08,Qi08}. This property can also be exploited to identify correlated
topological insulators \cite{Ran08,Juricic12,Assaad13}. Dirac fermions on the
$\pi$ flux square lattice have been studied in \cite{Weeks10,Jia13}. 
Furthermore, twisted graphene multilayers have been identified as an instance of a two-dimensional TCI \cite{Kindermann13}.

The physics of the $\pi$KM model is surprisingly rich. In the noninteracting
case, and for each spin projection, it has Chern insulator \cite{Haldane88}
ground states characterized by Chern numbers $C=\pm 2$, separated by a
topological phase transition.  The band structure resembles that of the
nucleated topological phase in the Kitaev honeycomb lattice model
\cite{Kitaev06,Lahtinen11,Lahtinen12} which corresponds to the vortex sector
of the Kitaev model characterized by a $\pi$ flux vortex at each plaquette.

The spinful $\pi$KM model is found to have a trivial
$Z_{2}$ invariant. However, there exist two pairs of helical edge states
crossing at distinct points in the projected Brillouin zone, which are robust
with respect to single-particle scattering processes as long as translation
symmetry is preserved. An intriguing question, which we address in this manuscript using
bosonization and quantum Monte Carlo methods, is if the edge states are
robust to correlation effects. At half filling, we find that umklapp
scattering processes between the two pairs of edge states localize the edge
modes in the corresponding low-energy model, leading to a gap in the edge
states without breaking translation symmetry. This prediction is consistent
with quantum Monte Carlo results for the correlated edge states. Away from
half filling, umklapp scattering is not relevant, and the edge states remain
stable provided that translation symmetry is not broken by disorder. Finally,
we investigate the bulk phase diagram of the $\pi$KM model with an additional
Hubbard interaction. Our mean-field and quantum Monte Carlo results suggest
the existence of a magnetic phase transition that extends to weak coupling at
the quadratic band crossing point.

The paper is organized as follows. In Sec.~\ref{sec_model}, we introduce the
$\pi$KM model. Section~\ref{sec_qmc_method} provides a brief discussion of
the quantum Monte Carlo methods.  The bulk properties are discussed in
Sec.~\ref{sec_nonint_bulk} (noninteracting case) and Sec.~\ref{sec_qmc_bulk}
(interacting case). Sec.~\ref{sec_nonint_edge} contains a discussion of the noninteracting edge states.
 The bosonization analysis of the edge states is
presented in Sec.~\ref{sec_boson}, followed by the quantum Monte Carlo
results for correlation effects on the edge states in
Sec.~\ref{sec_qmc_ribbon}.  Finally, we conclude in Sec.~\ref{sec_sum}.
\section{$\pi$ Kane-Mele-Hubbard model}\label{sec_model}
The KM model describes electrons on the honeycomb lattice with
nearest-neighbor hopping and spin-orbit coupling \cite{KaneMele05a}. Given the $U(1)$ spin
symmetry which conserves the ${z}$ component of spin, the KM Hamiltonian
reduces to two copies of the Haldane model \cite{Haldane88,Wright13}, one for each
spin sector.  The latter has an integer quantum Hall ground state or, in
other words, it is a Chern insulator.  The quantum spin Hall insulator
results when the two Haldane models are combined in a way that restores
time-reversal symmetry.

Here, we construct a new model (referred to as the $\pi$KM model) by taking
the KM model and inserting a magnetic flux $\pm\pi$ into each hexagon of the
underlying honeycomb lattice. Each flux can be thought of as originating from 
a time-reversal symmetry preserving magnetic field of the form
\begin{equation}
  \bm{B}_{\pm}(\bm{r})=
  \pi
  \delta(
  \bm{r}-\bm{r_{i}}
  )
  (\pm)\bm{e}_{z}\;,
\end{equation}
 and is given by
\begin{equation}
  \phi_{\pm} = \frac{h c}{e}\int_{\hexagon}\bm{B}_{\pm}(\bm{r})d\bm{S}=\pm\pi \frac{h c}{e}.
\end{equation}
As illustrated in Fig.~\ref{fig_band}(a), such an arrangement of fluxes of
size $\pm\pi$ (in units of $hc/e$) leads to a model with a unit cell
consisting of two hexagons.

For each spin projection $\sigma$, the Hamiltonian takes the form of a
modified Haldane model \cite{Haldane88},
\begin{eqnarray}
  \label{eqn_CI}
  \mathcal{H}^{\sigma}&=&
  -\sum_{\langle\bm{i},\bm{j}\rangle}
  \left[t(\bm{i},\bm{j})-\mu \delta_{\bm{ij}}\right]
  \hat{c}_{\bm{i},\sigma}^{\dagger}\hat{c}_{\bm{j},\sigma}^{\phantom\dagger}\\
  &&\quad+{i\sigma \sum_{\langle\langle\bm{i},\bm{j}\rangle\rangle}} \lambda(\bm{i},\bm{j})
  \nu_{\bm{i},\bm{j}}\hat{c}_{\bm{i},\sigma}^{\dagger}\hat{c}_{\bm{j},\sigma}^{\phantom\dagger}\;.\nonumber
\end{eqnarray}
Here, $t(\bm{i},\bm{j})=t\tau_{\bm{i},\bm{j}}$ and
$\lambda(\bm{i},\bm{j})=\lambda\tau_{\bm{i},\bm{j}}$ are the
nearest-neighbor and next-nearest-neighbor hopping parameters, respectively;
$\bm{i},\bm{j}$ index both lattice and orbital sites and $\mu$ is the
chemical potential. The factor $\nu_{\bm{i},\bm{j}}$ is
$-1$ ($+1$) for $\bm{i},\bm{j}$ indexing the orbitals $1$
or $3$ ($2$ or $4$).

The additional, nonuniform hopping phase factors $\tau_{\bm{i},\bm{j}}=\pm1$
account for the presence of the $\pi$ fluxes. A $\pi$ flux is inserted in a
honeycomb plaquette by choosing the phase factors $\tau_{\bm{i},\bm{j}}$ in
such a way that their product along a closed contour around the plaquette is
\begin{equation}
  \label{eqn_tau}
  \tau_{\bm{i},\bm{j}}\tau_{\bm{j},\bm{k}}\cdots\tau_{\bm{l},\bm{i}} = -1\,.
\end{equation}
In a periodic system, $\pi$ fluxes can only be inserted in pairs. Each
hopping process from $\bm{i}$ to $\bm{j}$ that crosses the connecting line of
a flux pair acquires a phase $\tau_{\bm{i},\bm{j}}=-1$, which fixes the
position of both fluxes according to Eq.~(\ref{eqn_tau}). In general, there
is no one-to-one correspondence between the flux positions and the set of
$\tau_{\bm{i},\bm{j}}$, \ie, one eventually has to make a gauge choice. Due
to the geometry of the four-orbital unit cell, two gauges exist [see
Fig.~\ref{fig_band}(a)] which have unitarily equivalent Hamiltonians.

On a torus geometry, Hamiltonian~(\ref{eqn_CI}) becomes
\begin{equation}
  \mathcal{H}^{\sigma}=
  \sum_{\bm{k}} 
  c^{\dagger}_{\bm{k},\sigma}
  H^\sigma(\bm{k})
  c^{\phantom\dagger}_{\bm{k},\sigma}\,,
  \label{eqn_CI2}
\end{equation}
where $c^{\phantom\dagger}_{\bm{k},\sigma}
=(\hat{c}_{1,\bm{k},\sigma}^{\phantom\dagger},\hat{c}_{3,\bm{k},\sigma}^{\phantom\dagger},\hat{c}_{2,\bm{k},\sigma}^{\phantom\dagger},\hat{c}_{4,\bm{k},\sigma}^{\phantom\dagger})^{T}$ 
is the basis in which the nearest-neighbor term is block off-diagonal.
The Hamilton matrix $H^\sigma(\bm{k})$ can be expressed in terms of Dirac
$\Gamma$ matrices \cite{KaneMele05a},
$\Gamma^{(1,2,3,4,5)}=(\sigma_{x}\otimes \openone,\sigma_{z}\otimes \openone,
\sigma_{y}\otimes\sigma_{x}, \sigma_{y}\otimes\sigma_{y},
\sigma_{y}\otimes\sigma_{z})$ and their commutators
$\Gamma^{ab}=[\Gamma^{a},\Gamma^{b}]/(2i)$:
\begin{equation}\label{eqn_CI3}
  H^\sigma
  (\bm{k})=
  \mu\,\openone+
  \sum\limits_{a=1}^{5} d_{a}(\bm{k}) \Gamma^{a} + \sum\limits_{a<b=1}^{5} d^\sigma_{ab}(\bm{k}) \Gamma^{ab}\,.
\end{equation}
The nonvanishing coefficients $d_{a}(\bm{k})$ and $d^\sigma_{ab}(\bm{k})$ are
given in Table~\ref{tab_gamma}.
\begin{table*}
  \begin{ruledtabular}
    \begin{tabular}{l l l}
      $d_{1}(\bm{k})=-t\cos(\bm{k}\bm{a}_{2})$ & 
      $d^\sigma_{12}(\bm{k})=t\sin(\bm{k}\bm{a}_{2})$  & 
      $d^\sigma_{23}(\bm{k})=t \cos(\bm{k}\bm{a}_{1}/2)\cos(\bm{k}(\bm{a}_{1}/2-\bm{a}_{2}))$ 
      \\
      $d_{3}(\bm{k})=-\frac{t}{2}\big[\sin(\bm{k}\bm{a}_{2}) -\sin(\bm{k}(\bm{a}_{1}-\bm{a}_{2}))\big]$ & 
      $d^\sigma_{13}(\bm{k})=-2\sigma \lambda\sin(\bm{k}\bm{a}_{1}/2)\cos(\bm{k}\bm{a}_{1}/2)$ & 
      $d^\sigma_{24}(\bm{k})=\frac{t}{2}\big[\sin(\bm{k}(\bm{a}_{1}-\bm{a}_{2}))+\sin(\bm{k}\bm{a}_{2})\big]$
      \\
      $d_{4}(\bm{k})=-\frac{t}{2}\big[\cos(\bm{k}(\bm{a}_{1}-\bm{a}_{2}))-\cos(\bm{k}\bm{a}_{2})\big]$ & 
      $d^\sigma_{14}(\bm{k})=-2\sigma \lambda \sin^{2}(\bm{k}\bm{a}_{1}/2)$ & 
      $d^\sigma_{35}(\bm{k})=2\sigma \lambda\cos(\bm{k}\bm{a}_{1}/2)\cos(\bm{k}(\bm{a}_{1}/2-\bm{a}_{2}))$ 
      \\
      $d_{25}(\bm{k})=t$ &
      $d^\sigma_{15}(\bm{k})=2\sigma \lambda \sin(\bm{k}\bm{a}_{2})$ & 
      $d^\sigma_{45}(\bm{k})=2\sigma \lambda \cos(\bm{k}(\bm{a}_{1}/2-\bm{a}_{2}))\sin(\bm{k}\bm{a}_{1}/2)$ 
    \end{tabular}
    \caption{Nonzero coefficients $d_{a}(\bm{k})$ and $d^\sigma_{ab}(\bm{k})$
      of Eq.~(\ref{eqn_CI3}).\label{tab_gamma}}
  \end{ruledtabular}
\end{table*}

As for the KM model, a spinful and time-reversal invariant Hamiltonian
results by combining $\mathcal{H}^{\uparrow}$ and $\mathcal{H}^{\downarrow}$; $\lambda$ then
plays the role of an intrinsic spin-orbit coupling.  Including a Rashba
spin-orbit interaction which breaks the $U(1)$ spin symmetry, we have
\begin{eqnarray}
  \label{eqn_KM}
  \mathcal{H}_{0}&=&-\sum_{\langle\bm{i},\bm{j}\rangle,\sigma} 
  \left[t(\bm{i},\bm{j})-\mu\delta_{\bm{i},\bm{j}}\right]
  \hat{c}_{\bm{i},\sigma}^{\dagger}\hat{c}_{\bm{j},\sigma}^{\phantom\dagger}\\
  &&+{i\!\!\sum_{\langle\langle\bm{i},\bm{j}\rangle\rangle,\sigma}}\sigma\lambda(\bm{i},\bm{j})
  \nu_{\bm{i},\bm{j}}\hat{c}_{\bm{i},\sigma}^{\dagger}\hat{c}_{\bm{j},\sigma}^{\phantom\dagger}\nonumber\\
  &&+{i\sum_{\langle \bm{i},\bm{j}\rangle}}
  \left(\hat{c}^{\dagger}_{\bm{i},\uparrow}, \hat{c}^{\dagger}_{\bm{i},\downarrow}\right)
  \lambda_{\text{R}}(\bm{i},\bm{j})
  \bm{e}_{z} (\bm\sigma\times \bm{d}_{\bm{i},\bm{j}})
  \left(
    \begin{array}{c}
      \hat{c}^{\phantom\dagger}_{\bm{j},\uparrow} \\
      \hat{c}^{\phantom\dagger}_{\bm{j},\downarrow}
    \end{array}
  \right)\nonumber\;.
\end{eqnarray}
In the Rashba term,
$\lambda_{\text{R}}(\bm{i},\bm{j})=\lambda_{\text{R}}\tau_{\bm{i}\bm{j}}$,
$\bm{d}_{\bm{i},\bm{j}}$ is a vector pointing to one of the three
nearest-neighbor sites, and $\bm\sigma=(\sigma^x,\sigma^y,\sigma^z)$ is the
vector of Pauli matrices.

Taking into account a Hubbard term to model electron-electron interactions,
we finally arrive at the Hamiltonian of the $\pi$ Kane-Mele-Hubbard ($\pi$KMH) model,
\begin{equation}
  \label{eqn_KMH}
  \mathcal{H}=\mathcal{H}_{0}+U\sum_{\bm{i}}\hat{n}_{\bm{i},\uparrow} \hat{n}_{\bm{i},\downarrow}\,.
\end{equation}
\section{Quantum Monte Carlo methods}\label{sec_qmc_method}
The $\pi$KMH lattice model can be studied using the auxiliary-field determinant
quantum Monte Carlo method. Simulations are free of a sign problem given
particle-hole, time-reversal and $U(1)$ spin symmetry \cite{Hohenadler11,Zheng11, Hohenadler12}. 
This requirement excludes the $U(1)$ spin symmetry breaking Rashba term.
The algorithm has been discussed in detail previously \cite{Hohenadler12,AssaadBook08}.
To study the magnetic phase diagram of the $\pi$KMH model, we apply a
finite-temperature implementation \cite{AssaadBook08}.  The Trotter
discretization was chosen as $\Delta\tau t=0.1$. An inverse
temperature $\beta t=40$ was sufficient to obtain converged results.

Interaction effects on the helical edge states can be studied numerically by taking advantage 
of the exponential localization of the edge states and of the insulating nature of the bulk which has no low-energy excitations.
Accordingly, the low-energy physics is captured by considering the Hubbard term only for the edge sites at 
one edge of a (zigzag) ribbon.  The bulk therefore is considered
noninteracting and establishes the topological band structure; it plays the
role of a fermionic bath. The resulting model is simulated  
without further approximations using the continuous-time quantum Monte Carlo algorithm based on
a series expansion in the interaction $U$ (CT-INT) \cite{Rubtsov05}. 
A similar approach has previously been used to study edge
correlation effects in the KMH model \cite{Hohenadler11,Ho.As.11}. 
Compared to the KMH model, the Rashba term leads to a
moderate sign problem.
\section{Bulk properties of the $\pi$KM model}\label{sec_nonint_bulk}
In this section, we discuss the band structure and the topological phases of the noninteracting
model~(\ref{eqn_CI2}), corresponding to one spin sector of the $\pi$KM model.
 Subsequently, we show that
the spinful $\pi$KM model~(\ref{eqn_KM}) is $Z_{2}$ trivial at half filling.
\subsection{Band structure}
\begin{figure}[h]
  \includegraphics[width=0.5\textwidth]{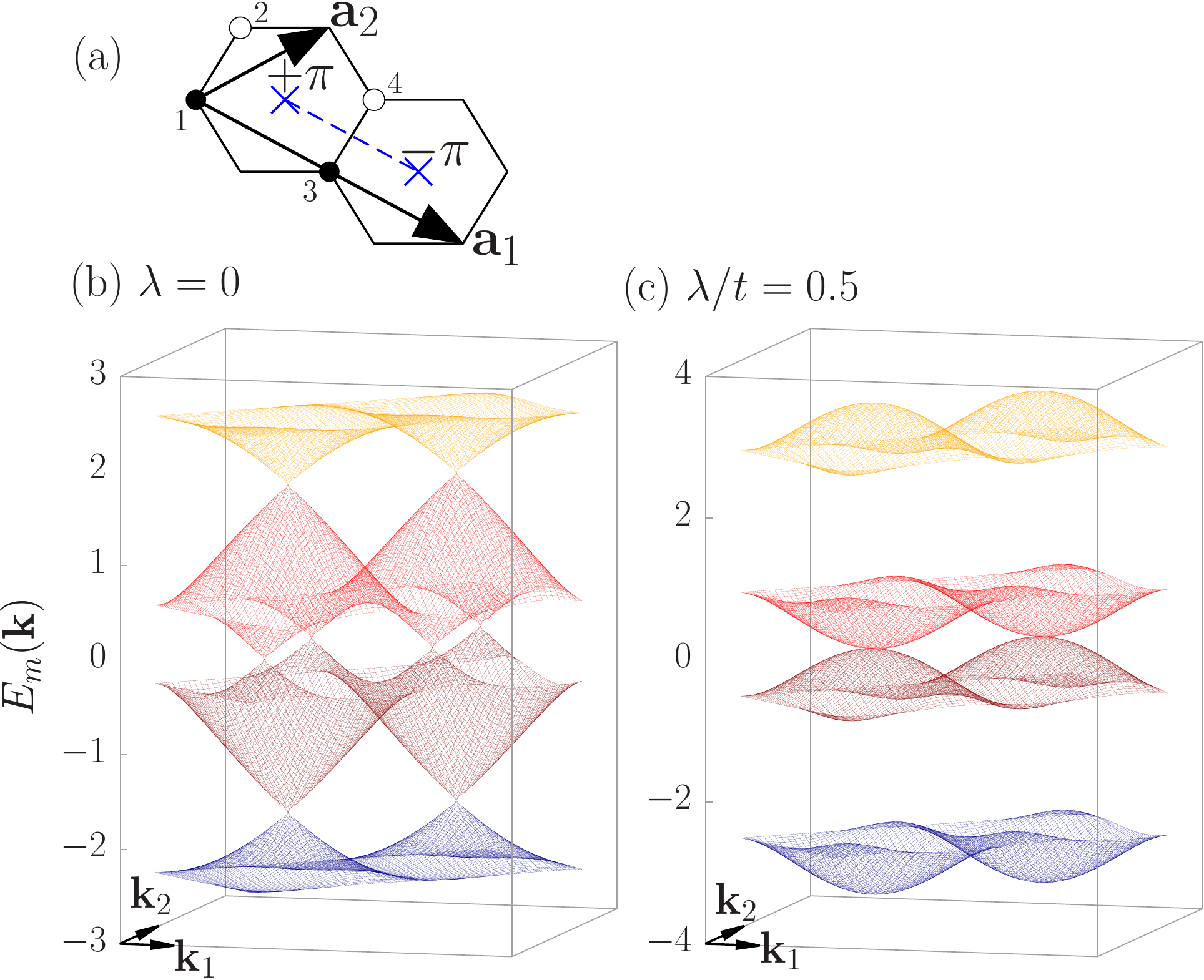}
  \caption{
(Color online) (a) The unit cell of the $\pi$ flux honeycomb lattice has four orbitals and is defined by the lattice vectors
 $\bm{a}_{1}=\big(3,-\sqrt{3}\big)$ and $\bm{a}_{2}=\frac{1}{2}\big(3,\sqrt{3}\big)$.
 Each honeycomb plaquette carries a magnetic flux $\pm\pi$.
 The flux positions, defined by Eq.~(\ref{eqn_tau}), are fixed by requiring that hopping terms crossing
the dashed blue line (which is a gauge choice) acquire a phase of $-1$. 
The eigenvalue spectrum $E_{m}(\bm{k})$  of $\mathcal{H}^{\sigma}$ [Eq.~(\ref{eqn_CI})] 
has (b) four Dirac cones at $\lambda=0$, and (c) two points of quadratic band
crossing at $\lambda/t=0.5$.\label{fig_band}}
\end{figure}

The band structure is established by the eigenvalues of Eq.~(\ref{eqn_CI2}) 
which are, for $\mu=0$, given by
\begin{eqnarray}
  E_{m}(\bm{k})&=&
  \pm \Big\{
  3 t^{2} + 6 \lambda^{2} - 2 \lambda^{2} f(\bm{k})\\
  &&\pm t \sqrt{
    2\big[ 3 \big( t^{2} + 8 \lambda^{2}\big) + \big( t^{2} -16 \lambda^{2} \big) f(\bm{k}) \big]
  }\;\Big\}^{1/2}\;,\nonumber
\end{eqnarray}
where $f(\bm{k})=\cos(\bm{k}\bm{a}_{1}) + \cos(2\bm{k}\bm{a}_{ 2}) -
\cos[\bm{k}(2\bm{a}_{2}-\bm{a}_{1})]$.  At $\lambda=0$,
$\mathcal{H}^{\sigma}$ has four distinct Dirac points $\bm{K}_{i}$ with
linear dispersion at zero energy,
\begin{equation}
  E(\bm{K}_{i}+\bm{k})=\sqrt{\frac{3}{2}}t \left(k_{x}+k_{y}\right) +\mathcal{O}(k^{2})\;,
\end{equation}
where $\bm{K}_{1,2}=(\pi/3)(1,\pm 2/\sqrt{3})$,
$\bm{K}_{3}=(\pi/3)(2,5/\sqrt{3})$ and $\bm{K}_{4}=(\pi/3)(2,1/\sqrt{3})$.
At $\lambda/t=1/2$, the spectral gap closes quadratically at two points
$\bm{\Gamma}_{i}$,
\begin{equation}
  E(\bm{\Gamma}_{i}+\bm{k})=\frac{3\sqrt{3}}{4} t \left(k_{x}^{2}+k_{y}^{2}\right) +\mathcal{O}(k^{4})\;,
\end{equation}
where $\bm{\Gamma}_{1}=(\pi/3)(1,0)$ and
$\bm{\Gamma}_{2}=(\pi/3)(2,\sqrt{3})$ (Fig.~\ref{fig_band}).  For the spinful
model (\ref{eqn_KM}) with nonzero Rashba coupling, the point of quadratic
band crossing is replaced by a finite region with zero band gap.
\subsection{Quantized Hall conductivity}
\begin{figure}
  \includegraphics[width=0.5\textwidth]{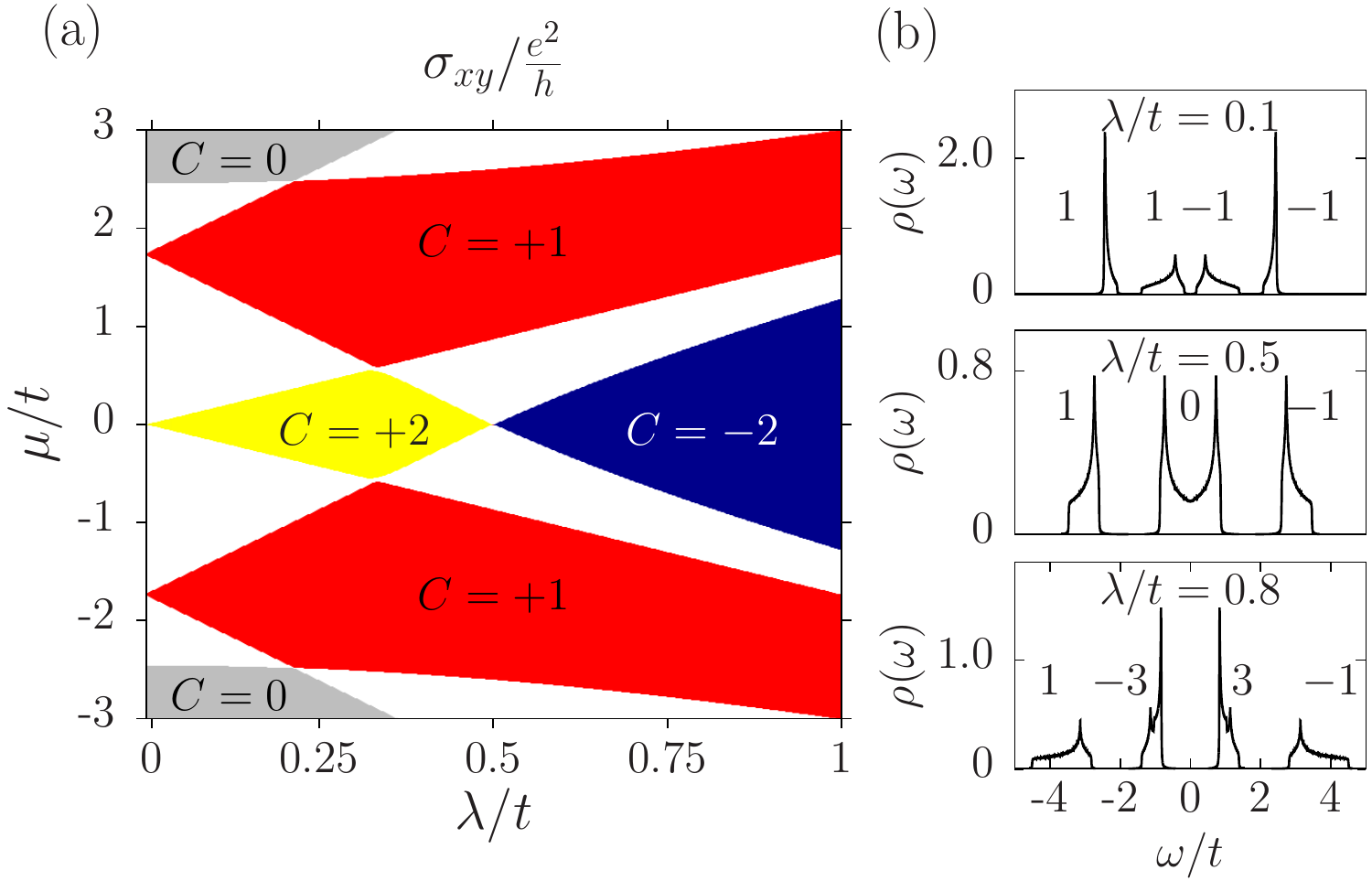}
  \caption{(Color online) (a) Total Chern number $C=\sum_{n}C_{n}$ of the
    occupied bands of $\mathcal{H}^{\downarrow}$ [Eq.~(\ref{eqn_CI})], as
    obtained from the Hall conductivity $\sigma_{xy}$ in the insulating phases which are separated by metallic regions (white).
    (b) Density of states $\rho(\omega)=(1/4N)\sum_{\bm{k},n}\delta(\omega-E_{n}(\bm{k}))$ and Chern numbers $C_{n}$ of the individual bands.
    \label{fig_map_sigma_xy_dos}}
\end{figure}

We first consider the Chern insulator defined by $H(\bm{k})$ in Eq.~(\ref{eqn_CI2}). 
In this case, the electromagnetic response reveals the topological properties of the band structure. 
In linear response
to an external vector potential, the optical conductivity tensor of an
$n$-band noninteracting system described by a Hamilton matrix $H(\bm{k})$ is
given by
\begin{equation}
  \sigma_{\alpha,\beta}(\omega)
  =
  \frac{1}{N}\frac{(e/\hbar)^{2}}{i(\omega +i0^{+})}
  \left[
    \langle K_{\alpha}\rangle \delta_{\alpha,\beta} - \Lambda_{\alpha,\beta}(\omega)
  \right]\;,
\end{equation}
where
\begin{eqnarray}
  \langle K_{\alpha}\rangle
  &=&
  \sum\limits_{\bm{k},n}
  f[E_{n}(\bm{k})]\text{Tr}[K_{\alpha}(\bm{k})P_{n}(\bm{k})]\;,\nonumber\\
  \Lambda_{\alpha,\beta}(\omega)
  &=&
  \sum\limits_{\bm{k},m,n}
  \lambda_{mn}(\bm{k},\omega)\text{Tr}[J_{\alpha}(\bm{k})P_{n}(\bm{k})J_{\beta}(\bm{k})P_{m}(\bm{k})]\;,\nonumber\\
  \lambda_{mn}(\bm{k},\omega)
  &=&
  \frac{f[E_{m}(\bm{k})]-f[E_{n}(\bm{k})]}{\omega + i0^{+} + E_{m}(\bm{k})-E_{n}(\bm{k}) }\;,
\end{eqnarray}
using the matrices $J_{\alpha}(\bm{k})=\partial H(\bm{k})/\partial
k_{\alpha}$, $K_{\alpha}(\bm{k})=-\partial^{2}H(\bm{k})/\partial
k_{\alpha}^{2}$, the projector on the $n$-th band $P_{n}(\bm{k})$, and the
Fermi function $f[E_{n}(\bm{k})]$. The Hall conductivity is then computed by
taking the zero-frequency limit of the optical conductivity,
\begin{equation}
  \underset{\omega\rightarrow 0}{\text{lim}} \operatorname{Re}\big[ \sigma_{xy}(\omega)\big]
  =\sigma_{xy}
  =\left[\sum\limits_{n=1}^{N_{\text{occ}}}C_{n}\right]\frac{e^{2}}{h}\,.
\end{equation}
It directly measures the (first) Chern number $C$ of the gap, which is the sum of the
Chern numbers $C_{n}$ of the $N_{\text{occ}}$ occupied bands.
Figure~\ref{fig_map_sigma_xy_dos} shows the Chern number as a function of the
chemical potential $\mu$ and the ratio $t_2/t_1$.  Transitions between
different Chern insulators are topological phase transitions and necessarily
involve an intermediate metallic state where the Chern number can in
principle take any value.  Of particular interest for the understanding of
correlation-induced instabilities is the transition at $\mu = 0$ as a
function of $t_2/t_1$ between the states with $C= \pm 2 $. At $t_2/t_1 =
1/2$, we find a a quadratic band crossing point with a nonzero density of
states.
 
For the spinful model~(\ref{eqn_KM}) with $U=0$ and a $U(1)$ spin symmetry
($\lambda_\text{R}=0$), one can define a quantized spin Hall conductivity
$\sigma^{s}_{xy}$ in terms of the Hall conductivity $\sigma^{\sigma}_{xy}$ of $\mathcal{H}^{\sigma}$ (\ref{eqn_CI2}).  At
$\mu=0$, $\sigma^{\sigma}_{xy}$ and $\sigma^{s}_{xy}$ take the values
\begin{equation}
  \label{eqn_sigmaxy}
  \sigma^{\sigma}_{xy}=\mp\sigma 2 \frac{e^{2}}{h}\;,\;\;\sigma^{s}_{xy}=\frac{\hbar}{2e}\left(\sigma^{\uparrow}_{xy}-\sigma^{\downarrow}_{xy}\right) =\mp 2 \frac{e}{2\pi}\;.
\end{equation}
The sign change occurs at the quadratic band crossing point at
$\lambda/t=\lambda_{0}=1/2$.
\subsection{$Z_{2}$ invariant}
In the general case where the $U(1)$ spin symmetry is broken, for example by
the presence of a Rashba term, the topological properties of a system with time-reversal
symmetry are determined by the $Z_2$ topological invariant \cite{KaneMele05b}. Recently, it
was shown that the $Z_2$ index can be calculated with a manifestly
gauge-independent method that only relies on time-reversal symmetry
\cite{Prodan11,Leung12}. The idea is to consider the adiabatic change of one
component of the reciprocal lattice vector, say $k_{y}$, along high-symmetry
paths $k_{y}\in(k,k^{\prime})$ in a rectangular Brillouin zone, while keeping
the other component ($k_{x}$) fixed.  This process is determined by the
unitary evolution operator $U_{k,k^{\prime}}$ and its differential equation
\begin{equation}
  \label{eqn_diffeq}
  i\frac{\text{d}}{\text{d}k}U_{k,k^{\prime}} = i\left[P_{k},\partial_{k} P_{k} \right]U_{k,k^{\prime}}\;.
\end{equation}
The initial condition is $U_{k^{\prime},k^{\prime}} =P_{k^{\prime}}$ and
$P_{k}=\sum_{i} |u_{i}(k)\rangle\langle u_{i}(k)|$ is the projector on the
occupied eigenstates of the $\pi$KM Hamiltonian. Equation~(\ref{eqn_diffeq}) is
integrated by evenly discretizing the path $(k,k^{\prime})$,
\begin{equation}
  U_{k,k^{\prime}}=\underset{N\rightarrow\infty}{\mathrm{lim}}\prod\limits_{n=1}^{N}P_{k^{\prime}+k\frac{n-1}{N-1}}\,.
\end{equation}
The topological invariant is then given as the product of two
pseudo-invariants
\begin{eqnarray}
  \Xi_{\text{2D}}=\pm 1&=&
  \prod\limits_{k_{x}=0,\pi}\frac{\text{Pf}\left[
      \langle u_{i}(0)|
      \theta
      |u_{j}(0)\rangle
    \right]}{\text{Pf}\left[
      \langle u_{i}(\pi)|
      \theta
      |u_{j}(\pi)\rangle
    \right]}\\\nonumber
  &&\quad\times\frac{
    \text{det}\left[
      \langle u_{i}(\pi)|
      U_{(\pi,0)} 
      |u_{j}(0)\rangle
    \right ]}
  {\sqrt{\text{det}\left[
        \langle u_{i}(\pi)|
        U_{(\pi,-\pi)}
        |u_{j}(\pi)\rangle
      \right ]}}\;,
\end{eqnarray}
where the dependence on $k_{x}$ is implicit and the invariant is computed
numerically \cite{Wimmer12}.  In the actual implementation, one has to make
sure to use the same branch for the square root at $k_{x}=0$ and at
$k_{x}=\pi$.  For the $\pi$KM  model~(\ref{eqn_KM}) at half filling ($\mu=0$) we
obtain, as expected \cite{Wu06}, a trivial insulator ($\Xi_{\text{2D}}=+1$).  In
contrast, if the chemical potential lies in the lower (upper) band gap,
{\ie}, at quarter (three-quarter) filling, we obtain a quantum spin Hall
insulator ($\Xi_{\text{2D}}=-1$).

It is interesting to consider how other bulk probes for the $Z_2$ index lead
to the conclusion of a trivial insulating state at half filling.  For
example, the $Z_{2}$ index can be probed by looking at the response to a
magnetic $\pi$ flux \cite{Qi08,Ran08,Assaad13}.  In the quantum spin Hall
state, threading a $\delta$-function $\pi$ flux through the lattice amounts
to generating a Kramers pair of states located at the middle of the
gap. Provided that the particle number is kept constant during the adiabatic
pumping of the $\pi$ flux, these mid-gap states give rise to a Curie law in
the uniform spin susceptibility.  This signature of the quantum spin Hall
state has been detected in Ref.~\onlinecite{Assaad13} in the presence of
correlations.  For the half-filled $\pi$KM model, the insertion of a $\pi$ flux
leads to a pair of Kramers degenerate states which form bonding and
antibonding combinations and thereby cut off the Curie law at energy scales
below the bonding-antibonding gap.
\section{Bulk correlation effects}\label{sec_qmc_bulk}
We begin our analysis of the effect of electron-electron interactions by
considering the $\pi$KMH model~(\ref{eqn_KMH}) on a torus geometry. In order to
compare our mean-field predictions to quantum Monte Carlo results, we set the
Rashba spin-orbit coupling and the chemical potential to zero.
\begin{figure}
  \includegraphics[width=0.5\textwidth]{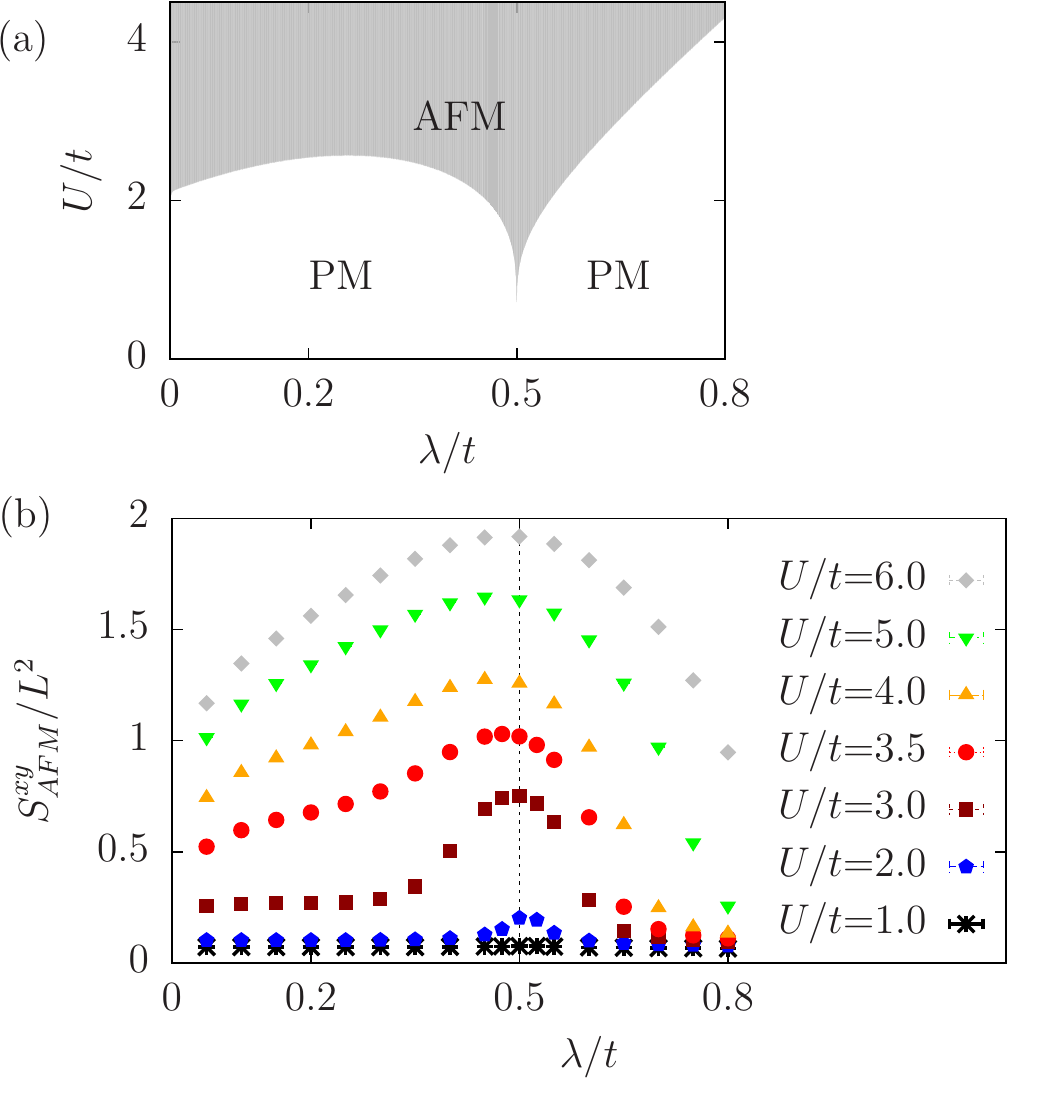}
  \caption{(Color online) (a) Phase diagram of the mean-field
    Hamiltonian~(\ref{eqn_Hmf}), showing the existence of a magnetically
    order phase with $xy$ magnetic order above a critical value $U_\text{c}$
    that depends on the spin-orbit coupling $\lambda$. For $\lambda/t=0.5$,
    where the model has a quadratic band crossing point, magnetic order
    exists for any nonzero value of $U_\text{c}$.  (b) Transverse magnetic
    structure factor $S^{xy}_{\text{AFM}}$ of the model~(\ref{eqn_KMH}) for
    different values $U/t$, as obtained from quantum Monte Carlo simulations
    of the $\pi$KMH model on a $6 \times 6$ lattice with periodic boundary conditions and at
    inverse temperature $\beta t=40$.  }
  \label{fig_ftqmc}
\end{figure}
The KMH model without additional $\pi$ fluxes is known to exhibit long-range,
transverse antiferromagnetic order at large values of $U/t$ \cite{Rachel10,
  Hohenadler11,Zheng11,Laubach13}. We therefore decouple the Hubbard term in
Eq.~(\ref{eqn_KMH}) in the spin sector, allowing for an explicit breaking of
time-reversal symmetry.  The mean-field Hamiltonian reads
\begin{equation}\label{eqn_Hmf}
  \mathcal{H}_{\text{mf}}=\mathcal{H}_{0}
  -\frac{2U}{3} \sum_{\bm{i}}\big(2\hat{S}_{\bm{i}}\langle\hat{S}_{\bm{i}}\rangle
  -\langle\hat{S}_{\bm{i}}\rangle^{2}\big) +\frac{UN}{2}\;,
\end{equation}
where $\mathcal{H}_{0}$ is given by Eq.~(\ref{eqn_KM}) with
$\lambda_{\text{R}}=0$, and
$\hat{S}_{\bm{i}}=(\hat{S}_{\bm{i}}^{x},\hat{S}_{\bm{i}}^{y},\hat{S}_{\bm{i}}^{z})$.
Assuming antiferromagnetic order, we make the ansatz
$\langle\hat{S}_{\bm{i}}\rangle=S_{\text{mf},\bm{i}}$ and
\begin{eqnarray}\label{eqn_mfparam}
S_{\text{mf},\bm{i}}^{x}&=&\nu_{\bm{i}}\,m\,,\;\; S_{\text{mf},\bm{i}}^{y,z}=0\,,\nonumber\\
S_{\text{mf},\bm{i}}^{x}&=&\frac{1}{Z}
\frac{1}{2}\sum\limits_{s,s^{\prime}}
\text{Tr}\left[ e^{-\beta \mathcal{H}_{\text{mf}}\{S_{\text{mf},\bm{i}}^{x} \}}
\hat{c}_{\bm{i},s}^{\dagger}\sigma_{x}\hat{c}_{\bm{i},s^{\prime}}^{\phantom\dagger}
\right],
\end{eqnarray}
where $\nu_{\bm{i}}=+1$ ($\nu_{\bm{i}}=-1$) if $\bm{i}$ indexes the orbitals $1,3$ ($2,4$).
Equation~(\ref{eqn_mfparam}) is solved self-consistently, resulting in the phase diagram shown in Fig.~\ref{fig_ftqmc}(a). 
We find a magnetic phase with transverse antiferromagnetic order above a critical value
of $U/t$ which depends on $\lambda/t$. In particular, at the
quadratic band crossing point ($\lambda_{0}=0.5$), the magnetic
transition occurs at infinitesimal values of $U/t$ as a result of the
Stoner instability associated with the nonvanishing density of states
at the Fermi level. Tuning the system away from the quadratic band crossing
point, the critical interaction increases. 

To go beyond the mean-field approximation, we apply the auxiliary-field quantum
Monte Carlo method discussed in Sec.~\ref{sec_qmc_method} to the $\pi$KMH model.
We calculate the transverse antiferromagnetic structure factor 
\begin{equation}
S^{xy}_{\text{AFM}} =
\frac{1}{L^{2}}
\sum_{\bm{i},\bm{j}}
 (-1)^{\nu_{\bm{i}} + \nu_{\bm{j}}} 
\langle \hat{S}^{+}_{\bm{i}}\hat{S}^{-}_{\bm{j}} + \hat{S}^{-}_{\bm{i}}\hat{S}^{+}_{\bm{j}}\rangle
\end{equation}
as a function of the interaction $U$ and  the spin-orbit coupling $\lambda$. 
Simulations were done on a $6\times6$ $\pi$-flux honeycomb lattice
(equivalent to $72$ honeycomb plaquettes).

As shown in Fig \ref{fig_ftqmc}(b), for small $U/t$, the structure
factor has a clear maximum close to $\lambda_{0}$, where the
weak-coupling magnetic instability is observed in mean-field theory. At
larger values of $U/t$, the maximum becomes less pronounced, and the
enhancement of $S^{xy}_{\text{AFM}}$ for all values of $\lambda/t$ is
compatible with the existence of a magnetic phase for all $\lambda/t$ at
large $U/t$. These numerical results seem to confirm the overall features of
the mean-field phase diagram. The numerical determination of the exact phase
boundaries from a systematic finite-size scaling is left for future work.
\section{Edge states of the $\pi$KM model}\label{sec_nonint_edge}
\begin{figure}
   \includegraphics[width=0.5\textwidth]{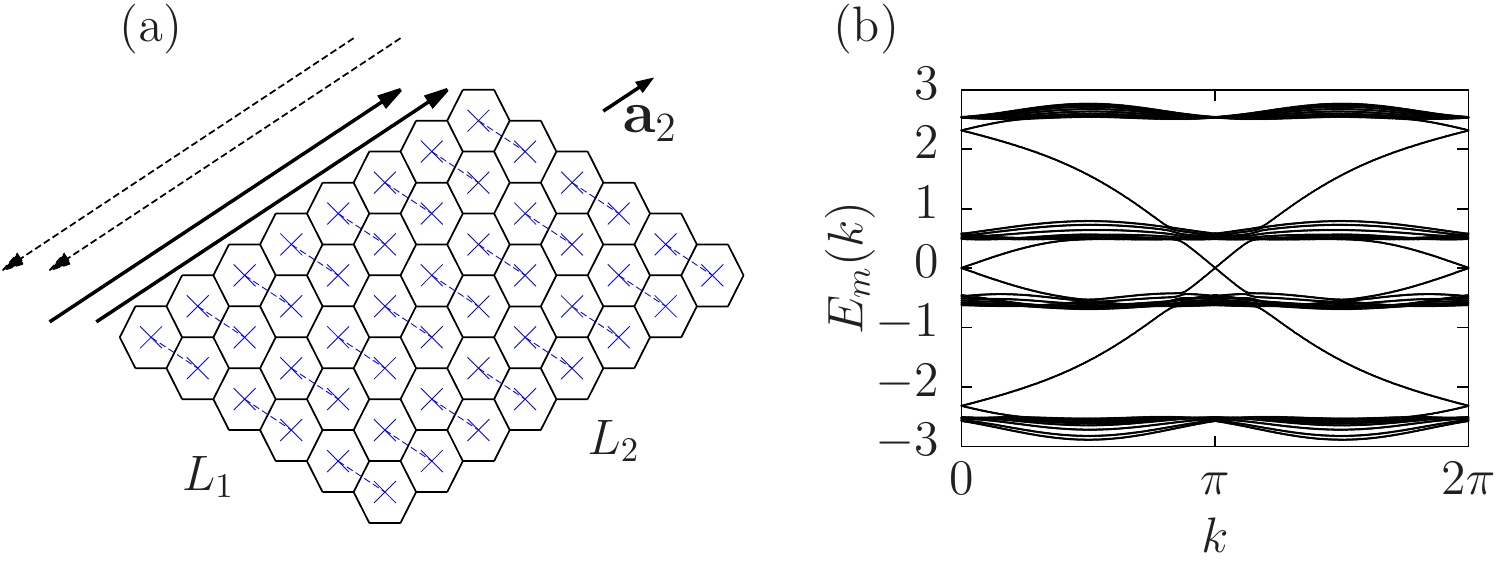}
   \caption{(Color online) (a) Ribbon geometry of the $\pi$ flux honeycomb
     lattice.  In the spinful case, the edge states consist of two Kramers
     doublets with Fermi velocities $v_{0}$ and $v_{\pi}$.  (b) Eigenvalue
     spectrum $E_{m}(k)$ of Eq.~(\ref{eqn_KM}) for $\lambda/t=0.3$ and
     $\lambda_{\text{R}}/t=0.1$ on a zigzag ribbon.\label{fig_band_ribbon2}}
 \end{figure}

We now consider the edge states of the noninteracting $\pi$KM model~(\ref{eqn_KM}) on a
zigzag ribbon with open (periodic) boundary conditions in the $\bm{a}_{1}$
($\bm{a}_{2}$) direction [Fig.~\ref{fig_band_ribbon2}(a)], and with momentum
$k=\bm{k}\cdot\bm{a}_{2}$ along the edge. Since the model is $Z_{2}$ trivial,
we expect an even number of edge modes to traverse the bulk gap \cite{Wu06}.  Furthermore,
given the spin Chern number $\sigma^{s}_{xy}/(e/2\pi)=\pm 2$ [see
Eq.~(\ref{eqn_sigmaxy})], we expect two helical edge modes at half filling.
Figure~\ref{fig_band_ribbon2}(b) shows the eigenvalue spectrum with
degenerate Kramers doublets at the time-reversal invariant momenta $k=0$ and
$k=\pi$. For $\lambda_{0}<\lambda/t<\lambda_{\pi}$, where
$\lambda_{\pi}=\sqrt{3}/2$, the eigenvalue spectrum of Eq.~(\ref{eqn_CI}) has
two additional cones at $k=\pi\pm\delta$. They are unstable in the sense that
their existence relies on the $U(1)$ spin symmetry.

The edge modes at $k=0$ ($k=\pi$) and $\sigma=\uparrow,\downarrow$ can be
further characterized by their Fermi velocity $v_{0}$ ($v_{\pi}$) and---in
the case of a $U(1)$ spin symmetry---by their chirality (the sign of the velocity).
The chirality changes at $\lambda_{0}$ and $\lambda_{\pi}$. 
For $\lambda/t<\lambda_{0}$, the edge modes have the same chirality,
so that the ($0,\sigma$) modes propagate in the same direction as the
($\pi,\sigma$) modes. In contrast, for $\lambda_{0}<\lambda/t<\lambda_{\pi}$, they have opposite
chirality since the  direction of propagation of the ($0,\sigma$) modes is reversed
after going through the point of quadratic band crossing.
At $\lambda/t=\lambda_{\pi}$, the additional cones at $k=\pi\pm\delta$ merge
with the  ($\pi,\sigma$) modes.  Consequently, the direction of propagation
of the  ($\pi,\sigma$) modes is reversed  and for $\lambda/t>\lambda_{\pi}$
both edge modes have the same chirality again. In the limit
$\lambda/t\rightarrow\infty$,  $v_{0}$ and $v_{\pi}$ become
equal. Furthermore, the velocities have equal magnitude but opposite sign at
$\lambda/t=\lambda_{s}\approx 0.665$.

\begin{figure}
  \includegraphics[width=0.5\textwidth]{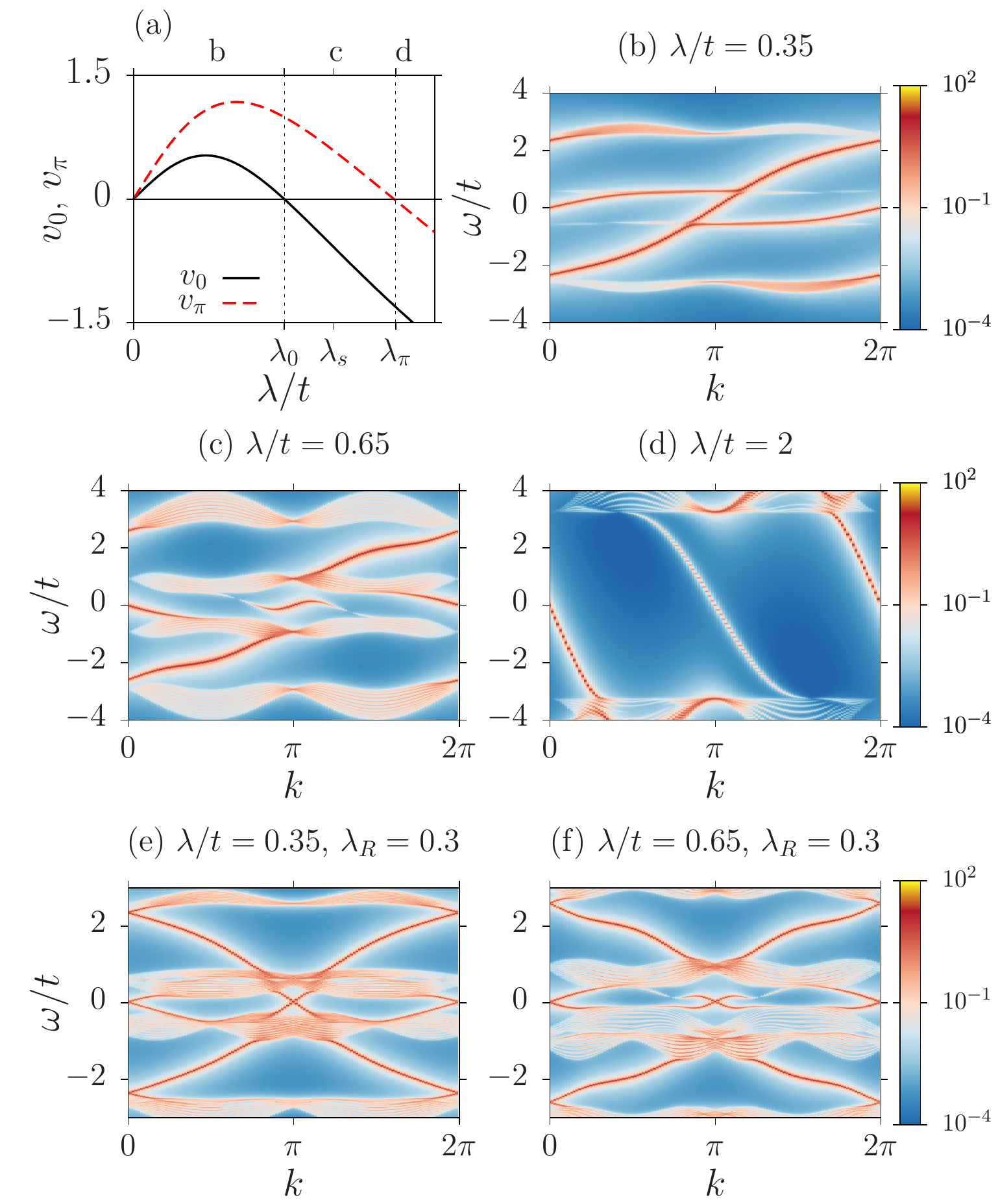}
  \caption{(Color online) (a) The Fermi velocity
    $v_{0}$ ($v_{\pi}$) changes sign at $\lambda_{0}$ ($\lambda_{\pi}$) so
    that for $\lambda_{0}<\lambda<\lambda_{\pi}$, the ($0,\sigma$) and
    ($\pi,\sigma$) edge modes have opposite chirality.  $\lambda_{s}$
    defines a symmetric point where $v_{0}=-v_{\pi}$ holds.  (b)--(d)
    Single-particle spectral function $A^{\uparrow}(k,\omega)$ along the
    edge.  (e),(f) Spin-averaged single-particle
    spectral function $A(k,\omega)=\sum_{\sigma}A^{\sigma}(k,\omega)/2$ along
    the edge. Here, $\lambda_\text{R}=0$ in (a)--(d), and
    $\lambda_\text{R}/t=0.3$ in (e),(f). \label{fig_aom_edge3}}
\end{figure}

To study the edge states, we consider the local single-particle spectral function
\begin{equation}
\label{eqn_spectral}
A^{\sigma}_{i}(k,\omega)= -\frac{1}{\pi}\mathrm{Im}\;G_{ii}^{\sigma}(k,\omega+i0^{+})\;,
\end{equation}
where the local noninteracting Green function is
\begin{equation}
G_{ii}^{\sigma}(k,\omega+i0^{+})=\left[\omega+i0^{+}-H(k) \right]_{i\sigma,i\sigma}^{-1}\;.
\end{equation}
The edge corresponds to the  orbital index $i=2$ [Fig.~\ref{fig_band}(a)] and for brevity we will omit the index $i$ in the following.
The Fermi velocities $v_{0}$ and $v_{\pi}$ and the local spectral function are shown in Fig.~\ref{fig_aom_edge3} \footnote{The color schemes are based on gnuplot-colorbrewer; 10.5281/zenodo.10282.}. 

Similar phases, characterized by a trivial $Z_{2}$ index and  two helical edge modes at $k=0,\pi$,  have been found in the KM model with additional third-neighbor hopping terms \cite{Hung14}, 
and in the anisotropic Bernevig-Hughes-Zhang model \cite{Bernevig06,Jiang14}.

In the remainder of this section, we concentrate on the low-energy properties of the $\pi$KM model~(\ref{eqn_KM}). 
Furthermore, we focus on the edge modes at the time-reversal invariant momenta $k=0,\pi$, and neglect the 
two additional, unstable modes at $k=\pi\pm\delta$ occurring for $\lambda_{0}<\lambda/t<\lambda_{\pi}$ which are gapped out by any finite Rashba coupling. 
Then, the effective Hamiltonian can be written in terms of 
right (left) moving fields $R_{1}(x)$ [$L_{1}(x)$] at the Fermi wave vector  $k_{\text{F}}^{(1)}=0$ and right (left) moving fields $R_{2}(x)$ [$L_{2}(x)$] at 
$k_{\text{F}}^{(2)}=\pi$:
\begin{equation}
\label{eqn_effham_x}
\mathcal{H}=\int \mathrm{d}x\bm{\Psi}^{\dagger}(x) H_{\text{edge}}(-i\partial_{x}) \bm{\Psi}(x)\,,
\end{equation}
where $ \bm{\Psi}^{\dagger} (x)  = ( 
R_{1}^{\dagger} (x) ,
L_{1}^{\dagger} (x) ,
R_{2}^{\dagger} (x) ,
L_{2}^{\dagger} (x) )$.
The chiral fields  have the anticommutation relations
\begin{eqnarray}
\label{eqn_anticom}
\{R_{i}(x),R_{j}^{\dagger}(x^{\prime})\}&=&
\{L_{i}(x),L_{j}^{\dagger}(x^{\prime})\}
=
\delta_{ij}\delta(x-x^{\prime})\,,\nonumber\\
\{R_{i}(x),L_{j}^{\dagger}(x^{\prime})\}
&=&
\{L_{i}(x),R_{j}^{\dagger}(x^{\prime})\}=0\;.
\end{eqnarray}
In the $U(1)$ spin symmetric case, we have
\begin{equation}
H_{\text{edge}}(-i\partial_{x}) = -i \partial_{x}\;\text{diag}(v_{1},v_{2})\otimes \sigma_{z}\;.
\end{equation}
Hamiltonian~(\ref{eqn_effham_x}) will be the starting point for the
bosonization analysis in Sec.~\ref{sec_boson}. 
\subsection{Effective low-energy model}\label{subsec_effmodel}
The edge of a two-dimensional bulk has two time-reversal invariant momenta, $k=0$ and $k=\pi$, and therefore several possibilities exist
 to have two pairs of helical edge states:
(i) both Kramers doublets cross at $k=0$ (or $k=\pi$), 
(ii) one Kramers doublet crosses at $k=0$ while the other crosses at $k=\pi$, and 
(iii) each Kramers doublet has one branch at $-k$ (or $\pi-k$) and its
time-reversed branch at $+k$ (or $\pi+k$). 
In cases (i) and (iii), degenerate states  which
are not Kramers partners exist at the same momentum and can be mixed by single-particle
backscattering. The edge states (i) and (iii) are therefore unstable at the
single-particle level. In contrast, the edge states (ii) are stable
at the single-particle level if translation symmetry is preserved at the
edge, thereby forbidding scattering between states at $k=0$ and $k=\pi$.

The metallic edge modes of Eq.~(\ref{eqn_KM}) are an instance of case
(ii). Given time-reversal symmetry and no interactions, the edge states
remain gapless even in the generic case without $U(1)$ spin symmetry as long
as translation symmetry and hence the momentum $k$ along the edge is preserved. 
On the other hand, the states acquire a gap when time-reversal symmetry is
broken. This is the case in the presence of, for example, a Zeeman term that
also breaks the $U(1)$ spin symmetry.

To illustrate this point, we consider the most general time-reversal symmetric formulation of the model (\ref{eqn_effham_x}) in momentum space.
Let $R_{i}^{\dagger}(p)$ [$L_{i}^{\dagger}(p)$] create an electron with velocity  $v_{i}$ [$-v_{i}$] 
(where  $v_{1}\equiv v_{0}$ and $v_{2}\equiv v_{\pi}$) and momentum $k = p + (i-1) \pi$. 
Then, Eq.~(\ref{eqn_effham_x}) reads
\begin{equation}
\label{eqn_effham_p}
\mathcal{H}=\sum\limits_{p} \bm{\Psi}^{\dagger}(p) H_{\text{edge}}(p) \bm{\Psi}(p)\;,
\end{equation}
where
$ \bm{\Psi}^{\dagger} (p)  = ( 
R_{1}^{\dagger} (p) ,
L_{1}^{\dagger} (p) ,
R_{2}^{\dagger} (p) ,
L_{2}^{\dagger} (p) )$ and
\begin{equation}
\label{eqn_effham}
H_{\text{edge}}(p) = H_{\text{SO}}(p)  + H_{\text{S}}\;,
\end{equation}
where $H_{\text{SO}}(p)$ is a general spin-orbit term and $H_{\text{S}}$ a single-particle scattering term.
Time-reversal symmetry is preserved when
 $\Theta H_{\text{edge}}(p)\Theta^{-1}=H_{\text{edge}}(-p)$, where 
$\Theta = \Gamma^{3}\Gamma^{5} K$. Here, $K$ denotes  complex conjugation and
the $\Gamma$ matrices were defined in Sec.~\ref{sec_model}. 

The spin-orbit coupling 
\begin{equation}
H_{\text{SO}}=
p\left(
\begin{array}{cc}
  v_{1}   \bm{\sigma}\cdot\bm{e}_1    &   0\\
    0  &   v_{2}   \bm{\sigma}\cdot\bm{e}_2 
\end{array}
\right)=H_{U(1)}(p)+H_{\text{R}}(p)
\end{equation}
can be split into a $U(1)$ spin-symmetric term, $H_{U(1)}(p)$, and a Rashba term, $H_{\text{R}}(p)$. 
The (not necessarily equal) spin quantization axes are labeled by real unit vectors $\bm{e}_{i}$.
Choosing $\bm{e}_{i}$ to point along the $z$-axis one may write the $U(1)$ spin symmetric part as
\begin{equation}
H_{U(1)}(p)   
=  
p \left(
\begin{array}{cc}
  v_{1}^{\phantom{x}}   \sigma_{z} e_{1}^{z}   &     0 \\
   0     &   v_{2}^{\phantom{x}}   \sigma_{z} e_{2}^{z}
\end{array}
\right) = p\left( v_{+}\Gamma^{15} + v_{-} \Gamma^{34}\right)\,,
\end{equation}
where  $ v_{\pm} =  (  v_{1}^{\phantom{x}} e_{1}^{z} \pm v_{2}^{\phantom{x}} e_{2}^{z})/2 $. 
Note that the generator of the $U(1)$  spin symmetry is $\Gamma^{34}=\openone\otimes\sigma_{z}$.

One way to break the $U(1)$ spin symmetry is to include the Rashba term $H_{\text{R}}(p)$ by setting $\bm{e}_{1}\neq \bm{e}_{2}$. 
This can be accomplished by choosing, for example, $\bm{e}_{1}=(0,0,e_{1}^{z})^{T}$ and
$\bm{e}_{2}=(e_{2}^{x},e_{2}^{y},e_{2}^{z})^{T}$, leading to
\begin{eqnarray}
\label{eqn_rashba}
H_\text{R}(p)   
&=&
p v_{2} \left(
\begin{array}{cc}
   0    &     0 \\
   0     &   \sigma_{x} e_{2}^{x} + \sigma_{y} e_{2}^{y} 
\end{array}
\right)
\\\nonumber
&=& 
\frac{p v_{2}}{2}\left[(\Gamma^{45}-\Gamma^{13}) e_{2}^{x} - (\Gamma^{35}+\Gamma^{14})e_{2}^{y}\right]\,.
\end{eqnarray}
$H_{\text{S}}$ breaks the translation symmetry of the bulk model in the sense that it allows 
single-particle scattering between the $i = 1$ and $i=2$ branches of the
low-energy model. Its general, time-reversal symmetric form is
\begin{eqnarray}
H_{\text{S}}
&=&\left(
\begin{array}{cc}
  0 & h_{\text{S}}\\
  h_{\text{S}}^{\star}  & 0 
\end{array}
\right)
 =  \alpha_1 \Gamma^1 + \alpha_3 \Gamma^3 +  \alpha_4 \Gamma^4 +  \alpha_5\Gamma^5\nonumber\\
&=& H_{\text{S},U(1)} + H_{\text{S}^{\prime}}\;,
\end{eqnarray}
where $h_{\text{S}}$ denotes the corresponding complex $2\times 2$ matrix and $\alpha_i\in \mathbb{R}$. 
Note that $H_{\text{S}}$ generally breaks the $U(1)$ spin symmetry since
$[H_{\text{S}},\Gamma^{34}]=2i(\alpha_{4}\Gamma^{3}-\alpha_{3}\Gamma^{4})$.
Therefore, we write it as the sum of  a symmetry-preserving term, $H_{\text{S},U(1)}=\alpha_{1}\Gamma^{1} + \alpha_{5}\Gamma^{5}$, and 
a symmetry-breaking term, $H_{\text{S}^{\prime}}=\alpha_{3}\Gamma^{3} + \alpha_{4}\Gamma^{4}$. 

We consider the following three cases: (a) unbroken translation symmetry
and unbroken $U(1)$ spin symmetry, (b) broken translation symmetry but
unbroken spin symmetry, and (c) broken translation symmetry and broken spin symmetry.

\begin{figure}
  \includegraphics[width=0.5\textwidth]{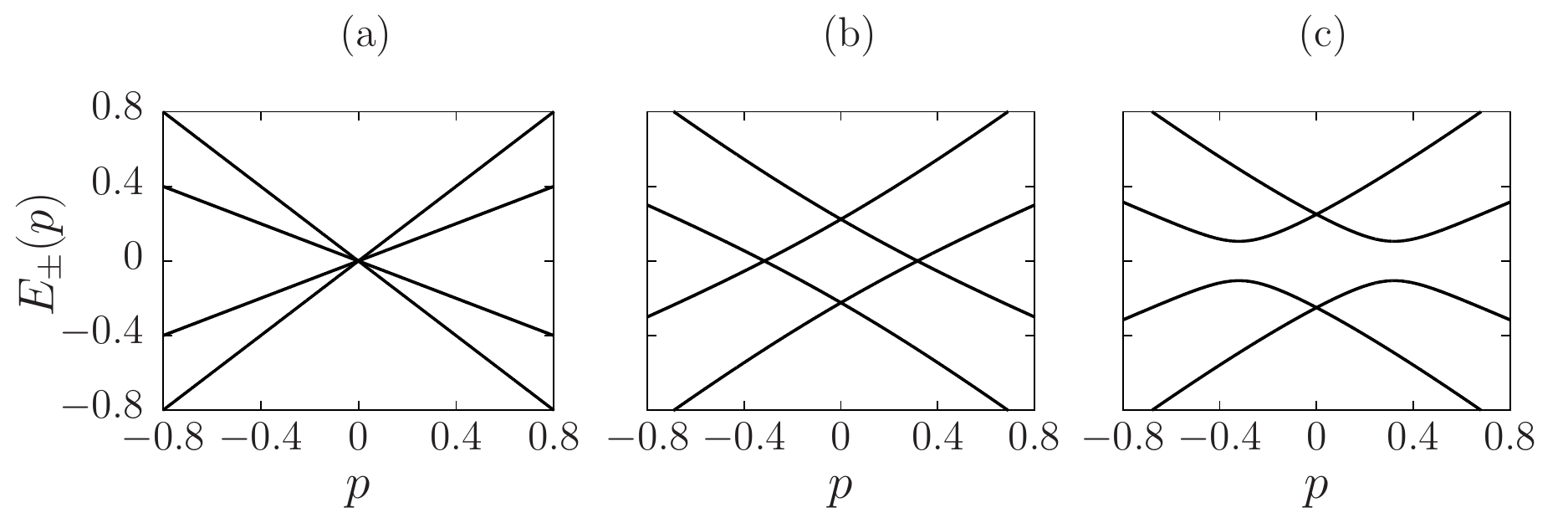}
  \caption{Spectrum $E_{\pm}(p)$ of the effective model
    (\ref{eqn_effham}), with $v_{1} =1$, $v_{2} = 0.5$, and $\bm{e}_1
    =\bm{e}_2 = \bm{e}_z$.  (a) Both translation symmetry and $U(1)$ spin
    symmetry are preserved ($\alpha_{i}=0$).  (b) Translation symmetry is
    broken, but $U(1)$ spin symmetry is preserved ($\alpha_1 = 0.2$,
    $\alpha_5 = 0.1$, $\alpha_3 =\alpha_4 =0 $).  (c) Both translation
    symmetry and $U(1)$ spin symmetry are broken ($\alpha_1 = 0.2$, $\alpha_5
    = 0.1$, $\alpha_3=0.1$, $\alpha_4 =0.05$). \label{fig_eff_edge}}
\end{figure}

In case (a), we have $H_{\text{S}}=0$, and $U(1)$ spin symmetry amounts to
$\bm{e}_{1}  = \bm{e}_{2}$.  This implies  $H_{\text{R}}(p)=0$, so that 
\begin{equation}
\label{eqn_effham_a}
H_{\text{edge}}^{(a)}(p)=H_{U(1)}(p)\;.
\end{equation}
The spectrum of $H_{\text{edge}}^{(a)}(p)$ is gapless, as shown in Fig.~\ref{fig_eff_edge}(a).

In case (b), we have
\begin{equation}
\label{eqn_effham_b}
H_{\text{edge}}^{(b)}(p)=H_{U(1)}(p) + H_{\text{S},U(1)}\,,
\end{equation}
and the spectrum, shown in Fig.~\ref{fig_eff_edge}(b), has two cones centered
at $p_{0}=\pm\sqrt{(\alpha_1^2 +  \alpha_5^2)/(v_+^2  -  v_-^2)}$, with the  
linearized dispersion 
\begin{equation}
E_{\pm}(p)=\pm \frac{v_+^2  -  v_-^2}{v_+}  (p \pm p_0) + \mathcal{O}(p^{2})\;.
\end{equation}
This illustrates that, as long as spin is conserved, the breaking of
translation symmetry does not gap out the edge states.

Finally, case (c) can be realized by adding the Rashba term
(\ref{eqn_rashba}) to Eq.~(\ref{eqn_effham_b}) or, alternatively, by 
considering 
\begin{equation}
\label{eqn_effham_c}
H_{\text{edge}}^{(c)}(p)=H_{U(1)}(p)+H_{\text{S}}\;,
\end{equation}
where $\alpha_{i}\neq 0$.  The resulting spectrum is gapped, see Fig.~\ref{fig_eff_edge}(c).

Returning to the original $\pi$KM model~(\ref{eqn_KM}), we expect the
combination of disorder (which breaks translation symmetry) and Rashba spin-orbit
coupling to open a gap in the edge states. 
We have measured the spin polarization carried by the helical edge modes 
as a function of disorder strength and using twisted boundary conditions \cite{Sheng06}.
Although the pair of Kramers doublets is in general not protected from localization by disorder, the 
spin polarization takes on finite values up to sizable disorder strengths. 
We attribute this finding to strong finite-size effects. The question of edge
state destruction by disorder deserves further investigation.
\subsection{Low-energy spin symmetries at $\lambda/t=\lambda_{s}$ and for $\lambda/t\rightarrow \infty$}\label{subsec_symmetry}
\begin{figure}
 \subfigure[]{
  \includegraphics[width=0.225\textwidth]{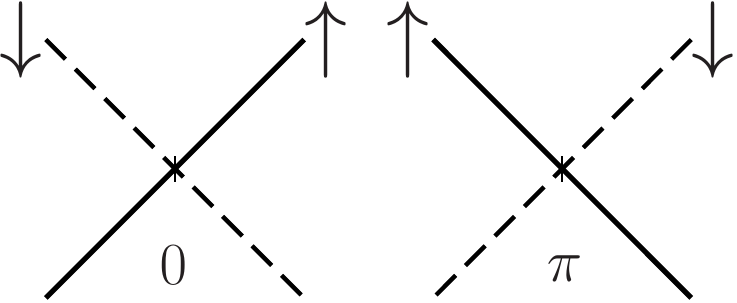}
  \label{fig_oppos_v}
  }
 \subfigure[]{
  \includegraphics[width=0.225\textwidth]{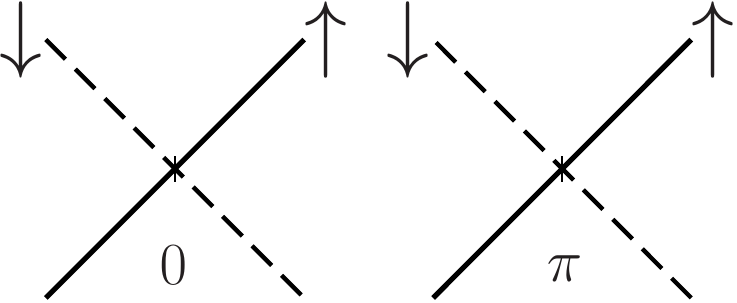}
  \label{fig_equal_v}
  }
   \caption{The ($0,\sigma$) and ($\pi,\sigma$) edge modes at (a)
     $\lambda/t=\lambda_{s}$ where $v_{0,\sigma}=-v_{\pi,\sigma}$, (b)
     $\lambda/t\rightarrow\infty$ where $v_{0,\sigma}=v_{\pi,\sigma}$.
 \label{fig_edge_vel}}
\end{figure}
In the following, we focus on two values of the intrinsic spin-orbit coupling,
$\lambda/t=\lambda_{s}$ and $\lambda/t\rightarrow\infty$, 
where the velocities of the ($0,\sigma$) and the ($\pi,\sigma$) modes
obey $v_{0,\sigma}=-v_{\pi,\sigma}$ and  $v_{0,\sigma}=v_{\pi,\sigma}$,
respectively (see Fig.~\ref{fig_edge_vel}). The corresponding low-energy
Hamiltonians are
\begin{equation}
\label{eqn_effham_s}
H_{\text{edge}}^{s}(-i\partial_{x})=-i\partial_{x} v \left(
\begin{array}{cc}
    \sigma_{z}    &     0 \\
   0     &  - \sigma_{z} 
\end{array}
\right)
=-i\partial_{x} v \Gamma^{15}\;,
\end{equation}
where $\bm{\Psi}^{\dagger}_{s} (x)  = ( 
R_{1}^{\dagger} (x) ,
L_{1}^{\dagger} (x) ,
L_{2}^{\dagger} (x) ,
R_{2}^{\dagger} (x) )$, 
and 
\begin{equation}
\label{eqn_effham_infty}
H_{\text{edge}}^{\infty}(-i\partial_{x})=-i\partial_{x} v \left(
\begin{array}{cc}
    \sigma_{z}    &     0 \\
   0     &  \sigma_{z} 
\end{array}
\right)
=-i\partial_{x} v \Gamma^{34}\;,
\end{equation}
where $\bm{\Psi}^{\dagger}_{\infty} (x)  = ( 
R_{1}^{\dagger} (x) ,
L_{1}^{\dagger} (x) ,
R_{2}^{\dagger} (x) ,
L_{2}^{\dagger} (x) )$. 
While the $SU(2)$ spin symmetry is obviously broken, we show in the following 
that a chiral $SU(2)$ symmetry exists for $\lambda/t=\lambda_{s}$.

The electron annihilation operator $\hat{c}_{\sigma}(x)$ can be written in
terms of the fields $R_{i}(x)$ and $L_{i}(x)$ \cite{Senechal99}, 
\begin{eqnarray}
\label{eqn_cop}
\hat{c}_{\uparrow}(x) &=& \left[ R_{1}(x) e^{-i k_{\text{F}}^{(1)} x} + Y_{2}(x) e^{-i k_{\text{F}}^{(2)} x}\right]/\sqrt{2}\;,\nonumber\\
\hat{c}_{\downarrow}(x) &=& \left[ L_{1}(x) e^{-i k_{\text{F}}^{(1)} x} + \bar{Y}_{2}(x) e^{-i k_{\text{F}}^{(2)} x}\right]/\sqrt{2}\;,
\end{eqnarray}
where $k_{\text{F}}^{(1)}=0$, $k_{\text{F}}^{(2)}=\pi$. For
$\lambda/t=\lambda_{s}$, the $i=1$ and $i=2$ modes have opposite helicity, so
$Y_{2}(x)=L_{2}(x)$ and $\bar{Y}_{2}(x)=R_{2}(x)$.
For $\lambda/t\rightarrow\infty$, we have  $Y_{2}(x)=R_{2}(x)$ and $\bar{Y}_{2}(x)=L_{2}(x)$.
The fermionic anticommutation relations follow from Eq.~(\ref{eqn_anticom}).
The spin operators can be expressed for both cases as 
\begin{eqnarray}
\label{eqn_spinop}
\hat{S}^{a}(x) &=& 
\frac{1}{2}\sum\limits_{\sigma,\sigma^{\prime}}
\hat{c}_{\sigma}^{\dagger}(x)\sigma^{a}_{\sigma,\sigma^{\prime}}
\hat{c}_{\sigma^{\prime}}^{\phantom\dagger}(x)\nonumber\\
&=&
\frac{1}{4}\sum\limits_{\sigma,\sigma^{\prime}}\Psi^{\dagger}_{\sigma}(x)
s^{a}_{\sigma,\sigma^{\prime}}
\Psi^{\phantom\dagger}_{\sigma^{\prime}}(x)\;,
\end{eqnarray}
with the constraint of single occupancy, $\hat{c}^{\dagger}_{\uparrow}(x)\hat{c}^{\phantom\dagger}_{\uparrow}(x) +\hat{c}^{\dagger}_{\downarrow}(x)\hat{c}^{\phantom\dagger}_{\downarrow}(x)=1$.
The matrices $s^{a}$ are given by
\begin{eqnarray}
\label{eqn_spinmat}
s^{x}&=& \openone \otimes \sigma_{x} + \left(\sigma_{x}\otimes\sigma_{x}\right) e^{i\pi x} = \Gamma^{45}-\Gamma^{23}e^{i\pi x}\nonumber\,,\\
s^{y}&=& \openone \otimes \sigma_{y} + \left(\sigma_{x}\otimes\sigma_{y}\right) e^{i\pi x} = -\Gamma^{35}-\Gamma^{24}e^{i\pi x}\nonumber\,,\\
s^{z}&=& \openone \otimes \sigma_{z} + \left(\sigma_{x}\otimes\sigma_{z}\right) e^{i\pi x} = \Gamma^{34}-\Gamma^{25}e^{i\pi x}\;.
\end{eqnarray}
They have the commutation relation $[s^{a}/4,s^{b}/4] = i \epsilon^{abc} (s^{c}/4)$. 

Apart from the spin operators, Eq.~(\ref{eqn_spinop}), there are three
additional operators  which have the commutation relations of the $su(2)$ Lie
algebra.  These operators are represented by the matrices
\begin{equation}
\label{eqn_chiral}
\Sigma_{x}\equiv \Gamma^{23}\;,\;\;\Sigma_{y}\equiv \Gamma^{24}\;,\;\;\Sigma_{z}\equiv \Gamma^{34}\;,
\end{equation}
which appear in Eq.~(\ref{eqn_spinmat}) and satisfy
$[\Sigma_{a}/2,\Sigma_{b}/2] = i \epsilon^{abc} (\Sigma_{c}/2)$.  They are
related to the additional chiral degree of freedom which is introduced by
the edge mode `orbitals' taking the values $i=1,2$.  For 
$\lambda/t=\lambda_{s}$, all three generators $\Sigma_{a}$ are symmetries of
the low-energy Hamiltonian (\ref{eqn_effham_s}), \ie,
$[H_{\text{edge}}^{s},\Sigma_{a}]=0$, whereas for
$\lambda/t\rightarrow\infty$, this is only true for $\Sigma_{z}$.  Therefore,
and apart from the spin symmetry, a chiral $SU(2)$ symmetry is present for 
$\lambda/t=\lambda_{s}$ which turns into a chiral $U(1)$ symmetry for
$\lambda/t\rightarrow\infty$.

We define a rotation by $\pi/2$, described by
\begin{eqnarray}
\label{eqn_rot}
U_{a}=\text{exp}\left[-i(\pi/4)\Sigma_{a}\right]=(\openone -i\Sigma_{a})/\sqrt{2}\;.
\end{eqnarray}
Then, $U_{a}^{\dagger}\hat{S}^{b}(x)U_{a}=M_{ab}$ 
is the rotation by $\pi/2$ of the spin component $\hat{S}^{b}(x)$ around the $\bm{e}_{a}$ axis, where 
\begin{equation}
\label{eqn_rot2}
M=\left(
\begin{array}{c c c}
\hat{S}^{x}(x) & e^{i\pi x}\hat{S}^{z}(x) &  -e^{i\pi x}\hat{S}^{y}(x) \\
-e^{i\pi x}\hat{S}^{z}(x) & \hat{S}^{y}(x) & e^{i\pi x}\hat{S}^{x}(x) \\
-\hat{S}^{y}(x) & \hat{S}^{x}(x) & \hat{S}^{z}(x)
\end{array}
\right)\;.
\end{equation}
In particular, we obtain the relations
\begin{eqnarray}
\label{eqn_symrel}
U_{x}^{\dagger}\hat{S}^{z}(x)U_{x} &=& -e^{i\pi x} \hat{S}^{y}(x)\;,\nonumber\\
U_{y}^{\dagger}\hat{S}^{z}(x)U_{y} &=& e^{i\pi x} \hat{S}^{x}(x)\;,\nonumber\\
U_{z}^{\dagger}\hat{S}^{y}(x)U_{z} &=& \hat{S}^{x}(x)\;.
\end{eqnarray}
We now consider the static spin structure factor
\begin{equation}
\label{eqn_spinspin}
S^{a}(q)=\frac{1}{\sqrt{N}}\sum\limits_{x} e^{-iqx}\langle \hat{S}^{a}(x) \hat{S}^{a}(0)\rangle\;, 
\end{equation}
where the expectation value is defined with respect to the effective Hamiltonian (\ref{eqn_effham_x}).
Using the symmetry relations~(\ref{eqn_symrel}) we get
\begin{eqnarray}
\label{eqn_symrel2}
S^{z}(q) &=& S^{x}(q+\pi)\quad \text{for}\,\lambda/t=\lambda_{s}\;,\nonumber\\
S^{x}(q) &=& S^{y}(q) \quad \hspace*{1.9em}   \text{for}\,\lambda/t=\lambda_{s}\,\text{and}\,\lambda/t\rightarrow\infty \;.
\end{eqnarray}
Equation~(\ref{eqn_symrel2}) relates the longitudinal and transverse
components of the spin-spin correlation functions.  In
Sec.~\ref{sec_qmc_ribbon}, we numerically show that this low-energy symmetry
is preserved in the presence of interactions.  It is therefore an emergent
symmetry of the interacting $\pi$KMH model~(\ref{eqn_KMH}). However, because
the chiral spins [Eq.~(\ref{eqn_chiral})] do not commute with the Rashba term
[\eg, Eq.~(\ref{eqn_rashba})], this symmetry hinges on $U(1)$ spin symmetry.
\section{Bosonization for the edge states}\label{sec_boson}
At low energies, the edge states of the $\pi$KMH model~(\ref{eqn_KMH}) can be
described in terms of a two-component \cite{Tanaka09,Orignac11, Tada12, Chung14}
Tomonaga-Luttinger liquid \cite{Delft98,Senechal99}. The Tomonaga-Luttinger
liquid is the stable low-energy fixed point of gapless interacting systems in
one dimension \cite{Haldane81}. We consider the free Hamiltonian with two left and two right
movers, forward scattering within the $i=1$ and $i=2$ branches
(intra-forward scattering of strength $g_{f}^{(i)}$), and  between the branches (inter-forward scattering of strength
$g_{f}^{\prime}$).  We focus on the case of two pairs of edge modes crossing
at $k=0$ and $k=\pi$, respectively, since only those are protected by time-reversal symmetry. 
In the following, we show that at half filling umklapp scattering between the edge modes is a relevant
perturbation in the sense of the renormalization group (RG).  It can drive
the model away from the Luttinger liquid fixed point and open gaps in the
low-energy spectrum.

We consider the following kinetic and interaction terms, 
\begin{eqnarray}
\label{eqn_ham_int1}
\mathcal{H}&=&
\sum\limits_{i=1}^{2}\Big[ v_{i}\int \mathrm{d}x\left(L_{i}^{\dagger} (i\partial_{x}) L_{i}^{\phantom\dagger}
+ R_{i}^{\dagger} (-i\partial_{x}) R_{i}^{\phantom\dagger}\right)\nonumber\\
&&\quad+ g_{f}^{(i)}\int \mathrm{d}x\,\rho_{i}^2\Big] 
+ g_{f}^{\prime}\int \mathrm{d}x\,\rho_{1}\rho_{2}\;,
\end{eqnarray}
where  $L_{i}$ ($R_{i}$) are the left (right) moving fields, and  $\rho_{i}=R_{i}^{\dagger}R_{i}+L_{i}^{\dagger}L_{i}$ 
is the electronic density.

To bosonize the above Hamiltonian~(\ref{eqn_ham_int1}), we introduce the bosonic fields $\phi_{i}(x)$, 
with $\partial_{x}\phi_{i}=\pi\rho_{i}$,  and $\Pi_{i}=R_{i}^{\dagger}R_{i}-L_{i}^{\dagger}L_{i}$, where 
$\left[\phi_{i}(x),\Pi_{i^{\prime}}(x^{\prime}\right]=i\delta_{i,i^{\prime}}\delta(x-x^{\prime})$. 
We then have
\begin{eqnarray}
\label{eqn_boson_1}
\mathcal{H}&=&
\frac{1}{2\pi}\int \mathrm{d}x
\sum\limits_{i=1}^{2}
\left[ v_{i}\left(\pi\Pi_{i}\right)^2 +  v_{i} K_{i}^{-2}\left(\partial_{x}\phi_{i}\right)^{2}\right]\nonumber\\
&\quad&+
\frac{g_{f}^{\prime}}{\pi^{2}}\int \mathrm{d}x\,\partial_{x}\phi_{1}\partial_{x}\phi_{2}\nonumber\\
&=&
\frac{1}{2\pi}\int \text{d}x \left[
\pi^{2} \Pi^{T}M\Pi + \left(\partial_{x}\phi\right)^{T} N \partial_{x}\phi
\right] \;,
\end{eqnarray}
where $K_{i}=(1+2g_{f}^{(i)}/\pi v_{i})^{-1/2}$ is a dimensionless parameter.
In the last line,
 we defined $\Pi=(\Pi_{1},\Pi_{2})^{T}$, $\phi=(\phi_{1},\phi_{2})^{T}$, and 
\begin{equation}
M=
\left(
\begin{array}{cc}
v_{1} &  0  \\
0 & v_{2}
\end{array}
\right)\,,\,
N=\frac{1}{\pi}
\left(
\begin{array}{cc}
\pi v_{1}+2g_{f}^{(1)} &  g_{f}^{\prime}  \\
g_{f}^{\prime}  & \pi v_{2}+2g_{f}^{(2)}
\end{array}
\right)\,,
\end{equation}
using the notation of Orignac \textit{et~al.}~\cite{Orignac10,Orignac11}.
The off-diagonal elements in $M$ are zero, since there is no single-particle scattering from the $i=1$ to the $i=2$ cone.
Hamiltonian (\ref{eqn_boson_1}) is decoupled by rescaling the fields:
\begin{eqnarray}
\label{eqn_boson_2}
\mathcal{H}&=&
\frac{1}{2\pi}\int \text{d}x \left[
\pi^{2} \Pi^{\prime T}\Pi^{\prime} + \left(\partial_{x}\phi^{\prime}\right)^{T} M^{1/2} N M^{1/2} \partial_{x}\phi^{\prime}
\right]\nonumber\\
&=&
\frac{1}{2\pi}\int \text{d}x \left[
\pi^{2} \Pi^{\prime\prime T}\Pi^{\prime\prime} + \left(\partial_{x}\phi^{\prime\prime}\right)^{T} \Delta \partial_{x}\phi^{\prime\prime}
\right]\nonumber\\
&=&
\frac{1}{2\pi}\int \text{d}x \sum\limits_{i=1}^{2}\Delta_{ii}^{1/2}\left[
\pi^{2} \widetilde{\Pi}_{i}^{2} + \left(\partial_{x}\widetilde{\phi}_{i}\right)^{2}
\right]\,,
\end{eqnarray}
where
$\Pi^{\prime}=M^{1/2}\Pi$, $\phi^{\prime}=M^{-1/2}\phi$, $\Pi^{\prime\prime}=S^{-1}\Pi^{\prime}$, $\phi^{\prime\prime}=S^{-1}\phi^{\prime}$,
 $\widetilde{\Pi}=\Delta^{-1/4}\Pi^{\prime\prime}$, and $\widetilde{\phi}=\Delta^{1/4}\phi^{\prime\prime}$.
$\Delta$ is a diagonal matrix and $S$ a rotation, defined via $\Delta=S^{-1}M^{1/2} N M^{1/2}S$. 
Therefore, the linear transformation to the new bosonic fields $\widetilde\Pi$ and
$\widetilde\phi$ is $\Pi=M^{-1/2} S \Delta^{1/4}\widetilde{\Pi}\equiv P\widetilde{\Pi}$ and  $\phi=M^{1/2} S \Delta^{-1/4}\widetilde{\phi}\equiv Q\widetilde{\phi}$. 
The  canonical commutation relations are preserved, since 
\begin{eqnarray}
\left[\widetilde{\phi}_{i}(x),\widetilde{\Pi}_{i^{\prime}}(x^{\prime}\right]
&=&
\sum\limits_{k,k^{\prime}}
Q^{-1}_{i,k}
\left(P^{-1}\right)_{k^{\prime},i^{\prime}}^{T}
\left[\phi_{k}(x),\Pi_{k^{\prime}}(x^{\prime})\right]\nonumber\\
&=&
i\delta_{i,i^{\prime}}\delta(x-x^{\prime})\;.
\end{eqnarray}
We have 
\begin{equation}
Q=
\left(
\begin{array}{cc}
S_{11} v_{1}^{1/2}\Delta_{11}^{-1/4} &  S_{12} v_{1}^{1/2} \Delta_{22}^{-1/4}   \\
S_{21} v_{2}^{1/2}\Delta_{11}^{-1/4}  & S_{22} v_{2}^{1/2} \Delta_{22}^{-1/4} 
\end{array}
\right)\;,
\end{equation}
\begin{eqnarray}
\Delta_{ii}&=&\frac{v_{1}N_{11} + v_{2}N_{22}}{2}\nonumber\\
\quad&&\pm\left[\left(\frac{v_{1}N_{11} - v_{2}N_{22}}{2}\right)^{2} + v_{1}v_{2} N_{12}^{2}\right]^{1/2}\,,
\end{eqnarray}
and, for $g_{f}^{\prime}\neq 0$, 
\begin{equation}
S=
\left(
\begin{array}{cc}
\frac{\text{sgn}(g_{f}^{\prime})}{\sqrt{1+s_{1}}}  & \frac{\text{sgn}(g_{f}^{\prime})}{\sqrt{1+s_{2}}}  \\
\frac{\text{sgn}\left(\Delta_{11}-v_{1}N_{11}\right)}{\sqrt{1+s_{1}^{-1}}} &  \frac{\text{sgn}\left(\Delta_{22}-v_{1}N_{11}\right)}{\sqrt{1+s_{2}^{-1}}}
\end{array}
\right)\;,
\end{equation}
where $s_{i}=(\Delta_{ii}-N_{11}v_{1})^{2}/v_{1}v_{2}N_{12}^{2}$. For $g_{f}^{\prime}=0$, $S=\openone$.
\begin{figure}
\subfigure[\;$g_{u}^{1}$ and $g_{u}^{2}$]{
  \includegraphics[width=0.225\textwidth]{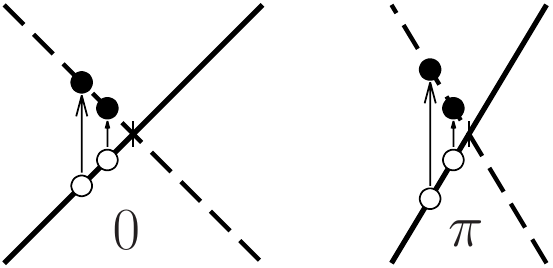}
  \label{fig_gu}
  }
\subfigure[\;$g_{u,1}^{\prime}$]{
  \includegraphics[width=0.225\textwidth]{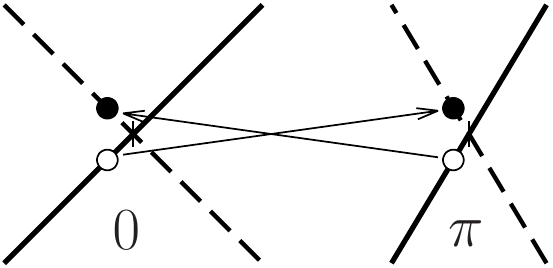}
  \label{fig_gup1}
  }
\subfigure[\;$g_{u,2}^{\prime}$]{
  \includegraphics[width=0.225\textwidth]{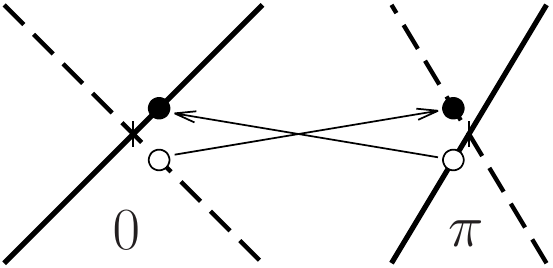}
  \label{fig_gup2}
  }
  \caption{The edge modes cross at $k=0$ and $k=\pi$ with in general nonequivalent Fermi velocities $v_{1}$ and $v_{2}$. 
   We consider the intra-umklapp scattering process (a), and the inter-umklapp
   scattering processes (b) and (c).
\label{fig_scattering}}
\end{figure}

We consider the following interactions as  perturbations to Eq.~(\ref{eqn_boson_2}):
intra-umklapp scattering of strength $g_{u}^{(i)}$ [Fig.~\ref{fig_gu}], inter-umklapp scattering of strength $g_{u,1}^{\prime}$ [Fig.~\ref{fig_gup1}],
 and inter-umklapp scattering of strength $g_{u,2}^{\prime}$
 [Fig.~\ref{fig_gup2}]. These processes are described by
\begin{eqnarray}
\mathcal{H}^{\prime}
&=&
\sum\limits_{i=1}^{2}g_{u}^{(i)}\int\mathrm{d}x\;
L_{i}^{\dagger}(x) L_{i}^{\dagger}(x+a) R_{i}^{\phantom\dagger}(x) R_{i}^{\phantom\dagger}(x+a)\nonumber\\
&\quad&\quad\times e^{i 4 k_{\text{F}}^{(i)}x}\nonumber\\
&\quad&+ g_{u,1}^{\prime}\int\mathrm{d}x\;
L_{1}^{\dagger}(x) L_{2}^{\dagger}(x) R_{1}^{\phantom\dagger}(x) R_{2}^{\phantom\dagger}(x) 
\;e^{i 2 (k_{\text{F}}^{(1)} +k_{\text{F}}^{(2)}) x}\nonumber\\
&\quad&+ g_{u,2}^{\prime}\int\mathrm{d}x\;
L_{1}^{\dagger}(x) R_{2}^{\dagger}(x) L_{2}^{\phantom\dagger}(x) R_{1}^{\phantom\dagger}(x) 
\;e^{i 2 (k_{\text{F}}^{(1)} - k_{\text{F}}^{(2)}) x}\nonumber\\
&\quad&+ \mathrm{H.c.}
\label{eqn_interact1}
\end{eqnarray}
The fermionic operators are
$R_{i}=\text{exp}(-i\phi_{R,i})/\sqrt{2\pi}$ and $L_{i}=\text{exp}(i\phi_{L,i})/\sqrt{2\pi}$, omitting the Klein factors, 
and we have  $\phi_{i}=(\phi_{R,i}+\phi_{L,i})/2$. 
We take $4k_{\text{F}}^{(i)}x=2 (k_{\text{F}}^{(1)} +k_{\text{F}}^{(2)})x=2 (k_{\text{F}}^{(1)}- k_{\text{F}}^{(2)})x=2\pi n$,
corresponding to half-filled bands. 
Then, 
\begin{eqnarray}
\mathcal{H}^{\prime}
&=&
\sum\limits_{i=1}^{2}\frac{g_{u}^{(i)}}{2\pi^{2}}\int\mathrm{d}x\;
\text{cos}\left(4\phi_{i}\right)
+ \frac{g_{u,1}^{\prime}}{2\pi^{2}}\int\mathrm{d}x\;
\text{cos}\left[2\left(\phi_{1}+\phi_{2}\right)\right]
\nonumber\\
&\quad&+ \frac{g_{u,2}^{\prime}}{2\pi^{2}}\int\mathrm{d}x\;\text{cos}\left[2\left(\phi_{1}-\phi_{2}\right)\right]\;.
\label{eqn_interact2}
\end{eqnarray}
We now consider $\mathcal{H}+\mathcal{H}^{\prime}$ and obtain the scaling dimensions $\Delta_{u}^{(i)}$, $\Delta_{u1}^{\prime}$, and $\Delta_{u2}^{\prime}$, 
of the vertex operators $\text{exp}(i4\phi_{i})$ and $\text{exp}[i2(\phi_{1}\pm\phi_{2})]$ in the above scattering processes \cite{Delft98}: 
\begin{eqnarray}
\Delta_{u}^{(1)}&=&4\left( Q_{11}^{2}+Q_{12}^{2}\right),\nonumber\\
\Delta_{u}^{(2)}&=&4\left( Q_{21}^{2}+Q_{22}^{2}\right),\nonumber\\
\Delta_{u1}^{\prime}&=&\left( Q_{11}+Q_{21}\right)^{2} + \left( Q_{12}+Q_{22}\right)^{2},\nonumber\\
\Delta_{u2}^{\prime}&=&\left( Q_{11}-Q_{21}\right)^{2} + \left( Q_{12}-Q_{22}\right)^{2}\;,
\end{eqnarray}
The scaling dimension $\Delta$ determines whether the respective
scattering process in $\mathcal{H}^{\prime}$ [Eq.~(\ref{eqn_interact2})] is a
relevant ($\Delta<2$) or irrelevant ($\Delta>2$) perturbation to the free
bosonic Hamiltonian $\mathcal{H}$ [Eq.~(\ref{eqn_boson_2})]. For
$g_{f}^{\prime}=0$, we have two separate Dirac cones, with $\Delta_{u}^{(i)}
= 4 v_{i}\Delta_{ii}^{-1/2}=4 K_{i}$ [see Eq.~(\ref{eqn_boson_1})]. 
Therefore, intra-umklapp scattering ($g_{f}^{(i)}$) becomes relevant when $K_{i}<1/2$, 
reproducing the result for a one-component helical liquid \cite{Wu06,Xu06}.

\begin{figure}
  \includegraphics[width=0.5\textwidth]{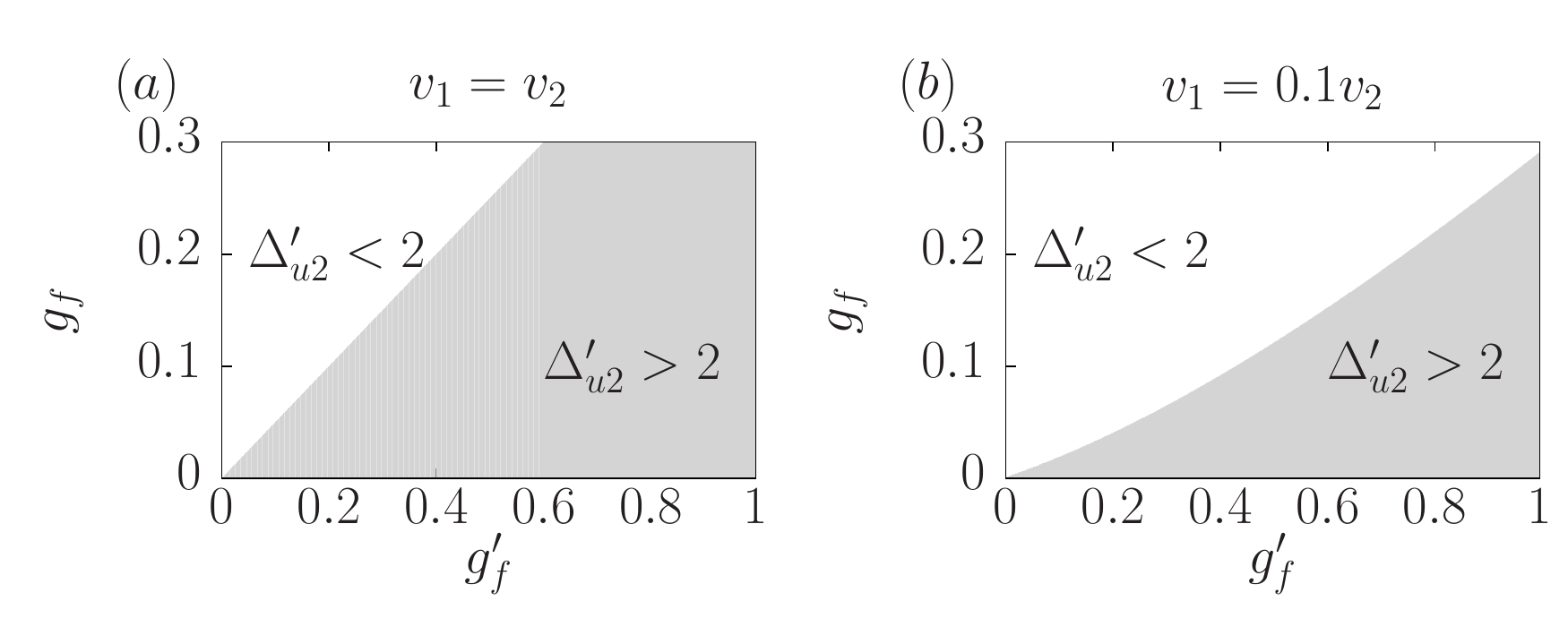}
  \caption{Phase diagram of the inter-umklapp process $g_{u,2}^{\prime}$ in the ($g_{f}^{\prime}$,$g_{f}$) plane
   for (a) equivalent  and (b) nonequivalent velocities of the edge
   modes. The scattering process is relevant (irrelevant) in the region where
   $\Delta_{u2}^{\prime}<2$ ($\Delta_{u2}^{\prime}>2$). 
   \label{fig_scalingdim}}
\end{figure}

In the case of weak coupling ($g_{f}^{1,2}\ll 1$ and $g_{f}^{\prime}\ll 1$),
we come to the following conclusions: (i) Intra-umklapp scattering is
RG-irrelevant, with $\Delta_{u}^{(1,2)}>2$. This is similar to the case of
the one-component helical liquid \cite{Wu06,Xu06}. (ii) Inter-umklapp
scattering $g_{u,1}^{\prime}$ is RG-relevant, with $\Delta_{u1}^{\prime}<2$.
(iii) The relevance of the inter-umklapp scattering $g_{u,2}^{\prime}$ is
determined by the phase diagram shown in Fig.~\ref{fig_scalingdim}.

If the $U(1)$ spin symmetry is preserved, only one of the two inter-umklapp
scattering processes $g_{u,1}^{\prime}$ or $g_{u,2}^{\prime}$
is allowed by symmetry, depending on the
chirality of the ($0,\sigma$) and ($\pi,\sigma$) modes which is determined by
the intrinsic spin-orbit coupling $\lambda$. As shown in
Fig.~\ref{fig_aom_edge3}(a), for $\lambda/t<\lambda_{0}$ and
$\lambda/t>\lambda_{\pi}$, both edge movers have the same chirality so that
inter-umklapp scattering corresponds to the $g_{u,2}^{\prime}$
term. In contrast, for $\lambda_{0}<\lambda/t<\lambda_{\pi}$, the edge movers have
opposite chirality and inter-umklapp scattering is given by the
$g_{u,1}^{\prime}$ term. 

The above distinction no longer holds when the $U(1)$ spin symmetry is
broken.  In this case, $g_{u,1}^{\prime}$ is always RG-relevant,
whereas the relevance of $g_{u,2}^{\prime}$ depends on the 
forward scattering strengths $g_{f}$ and $g_{f}^{\prime}$ and on the edge
velocities, see Fig.~\ref{fig_scalingdim}.

For $\lambda/t=\lambda_{\text{s}}$ ($\lambda/t\rightarrow\infty$), our
low-energy theory is similar to the fusion of two anti-parallel (parallel) helical edge modes
\cite{Tanaka09}, see also Fig.~\ref{fig_edge_vel}.  However, in the latter setup, the spatial overlap of the
two edge wave functions can be neglected, whereas it is included in the
interaction term of Eq.~(\ref{eqn_ham_int1}).
\section{Quantum Monte Carlo results for edge correlation effects}\label{sec_qmc_ribbon}
Correlation effects on the edge states of the $\pi$KMH model can be studied numerically using the
approach discussed in Sec.~\ref{sec_qmc_method}. Considering a zigzag ribbon,
we take into account a Hubbard interaction only at one edge, and simulate the
resulting model exactly using the CT-INT quantum Monte Carlo method.

We focus on two values of the spin-orbit coupling $\lambda/t$ and set the Rashba coupling to $\lambda_{\text{R}}/t=0.3$.  For $\lambda/t=0.35$, the
edge modes at $k=0$ and $k=\pi$ have different velocities ($v_{0}<v_{\pi}$),
whereas at $\lambda/t=0.65$, we have $v_{0}\approx v_{\pi}$. 
As in  the KMH model \cite{Ho.As.11},  we observe that the velocities of the edge states remain
almost unchanged with respect to the noninteracting case.

We carried out simulations for a zigzag ribbon of dimensions $L_{1}=25$ (open
boundary condition) and  $L_{2}=16$ (periodic boundary condition), see also
Fig.~\ref{fig_band_ribbon2}(a). For $\lambda_{\text{R}}=0$, $\mu=0$
corresponds to half filling. Although the band filling in general changes as
a function of $\lambda_{\text{R}}$ (the Rashba term breaks the particle-hole symmetry),
the Kramers degenerate edge states at $k=0,\pi$ are pinned to
$\omega=\mu$. The choice $\mu=0$ then again corresponds to half-filled Dirac
cones, and allows for umklapp scattering processes. The inverse
temperature was set to $\beta t=60$.
\begin{figure}[ht]
   \includegraphics[width=0.4\textwidth]{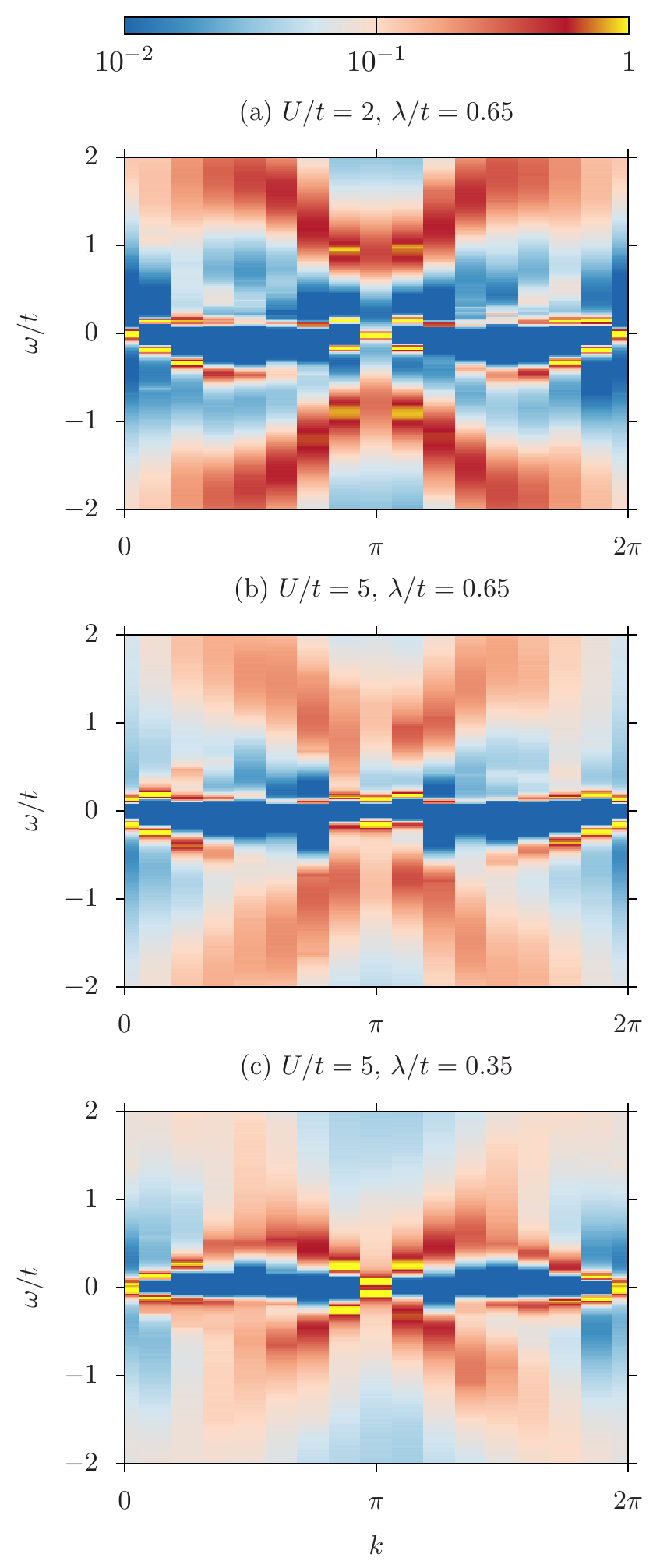}
   \caption{ \label{fig_ctqmc5} (Color online) Spin-averaged single-particle spectral function
     $A(k,\omega)$ [Eq.~(\ref{eqn_akwave})] from CT-INT simulations. (a)
     Weak coupling $U/t=2$, (b),(c) strong coupling $U/t=5$. Here,
     $\lambda_{\text{R}}/t=0.3$.} 
\end{figure}
\subsection{Single-particle spectral function}
Using CT-INT in combination with the stochastic maximum entropy method \cite{Beach04},
we calculate the spin-averaged spectral function at the edge,
\begin{eqnarray}\label{eqn_akwave}
  A(k,\omega) &=& \frac{1}{2}\sum_{\sigma} A^{\sigma}(k,\omega)\,,
  \\\nonumber
  A^{\sigma}(k,\omega)&=&-\frac{1}{\pi}\mathrm{Im}\;G^{\sigma}(k,\omega)\;,
\end{eqnarray}
where $G^{\sigma}(k,\omega)$ is the interacting single-particle Green
function, and $k$ is the momentum along the edge.  

As shown in Fig.~\ref{fig_ctqmc5}(a), for $U/t=2$, the
numerical results suggest the existence of gapless edge states. In contrast,
for a stronger interaction $U/t=5$, a gap is clearly visible both at $k=0$
and $k=\pi$. While the bosonization analysis in Sec.~\ref{sec_boson} predicts
a gap as a result of relevant umklapp scattering for any $U>0$, the size of
the gap depends exponentially on $U/t$. The apparent absence of a gap in
Fig.~\ref{fig_ctqmc5}(a) can therefore be attributed to the small system size
used ($L_2=16$).

Figure~\ref{fig_ctqmc5}(c) shows the spectral function~(\ref{eqn_akwave}) for 
$\lambda/t=0.35$, where $v_{0} < v_{\pi}$. Compared to the case of
$\lambda/t=0.65$ [Fig.~\ref{fig_ctqmc5}(b)] where $v_{0}\approx v_{\pi}$, the
gap in the edge states is much smaller. We expect this dependence on the
Fermi velocities to also emerge from the bosonization in the form of a
velocity-dependent prefactor that determines the energy scale of the gap
\cite{GiamarchiBook}.
\subsection{Charge and spin structure factors}
\begin{figure}
   \includegraphics[width=0.5\textwidth]{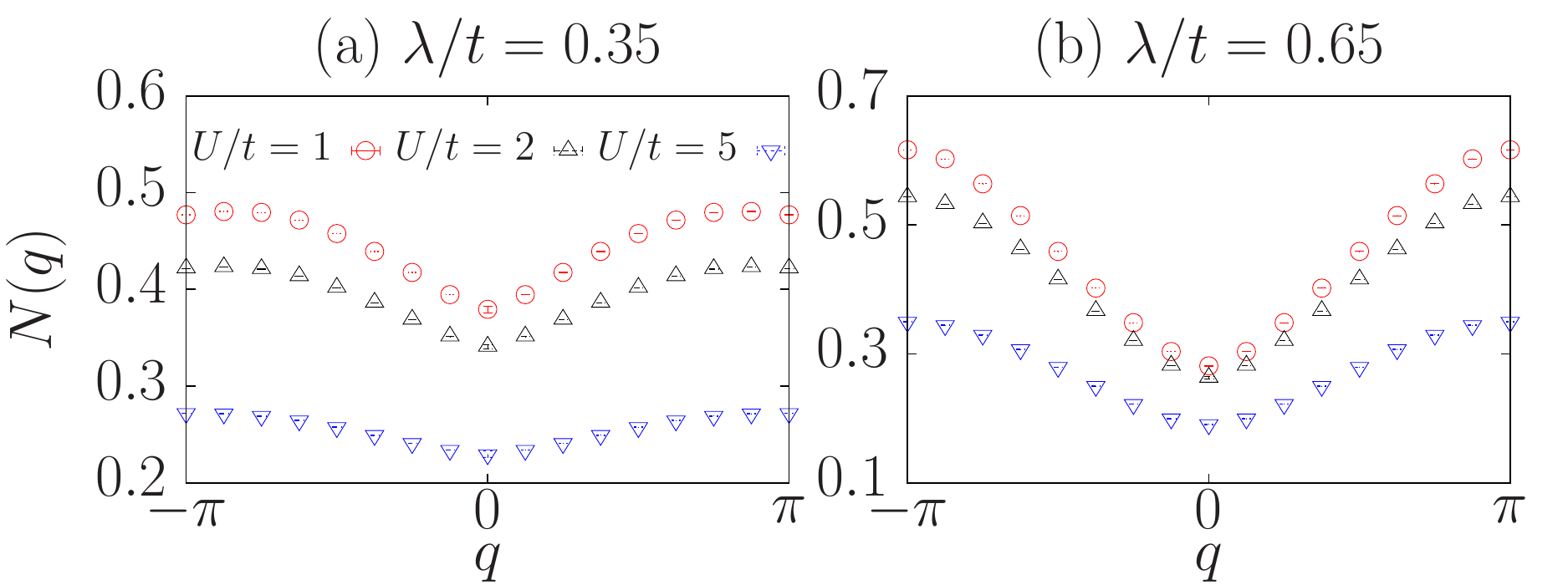}
  \caption{(Color online) Charge structure factor $N(q)$ [Eq.~(\ref{eq:nq})]
    from CT-INT simulations for (a) $\lambda/t=0.35$ and (b)
    $\lambda/t=0.65$. Here, $\lambda_{\text{R}}/t=0.3$. \label{fig_ctqmc7} } 
\end{figure}

We consider the charge structure factor
\begin{equation}\label{eq:nq}
N(q)=\frac{1}{\sqrt{N}}\sum\limits_{x} e^{-iqx}\left[ \langle \hat{n}(x) \hat{n}(0)\rangle -\langle \hat{n}(x)\rangle \langle\hat{n}(0)\rangle\right] \;,
\end{equation}
where $x$ is the position along the edge. 
Figure~\ref{fig_ctqmc7}(b) shows results for different values of $U/t$,
$\lambda/t=0.65$, and $\lambda_{\text{R}}/t=0.3$. For a weak interaction,
$U/t=1$, $N(q)$ exhibits cusps at $q=0$ and $q=\pi$ that indicate a power-law
decay of the real-space charge correlations. Upon increasing $U/t$, the cusps
becomes less pronounced, which suggests a suppression of charge correlations by the
interaction. This is in accordance with the existence of a gap in the single-particle
spectral function [Fig.~\ref{fig_ctqmc5}(b)]. A suppression of charge
correlations is also observed for $\lambda=0.35$, see Fig.~\ref{fig_ctqmc7}(a).

\begin{figure}
   \includegraphics[width=0.5\textwidth]{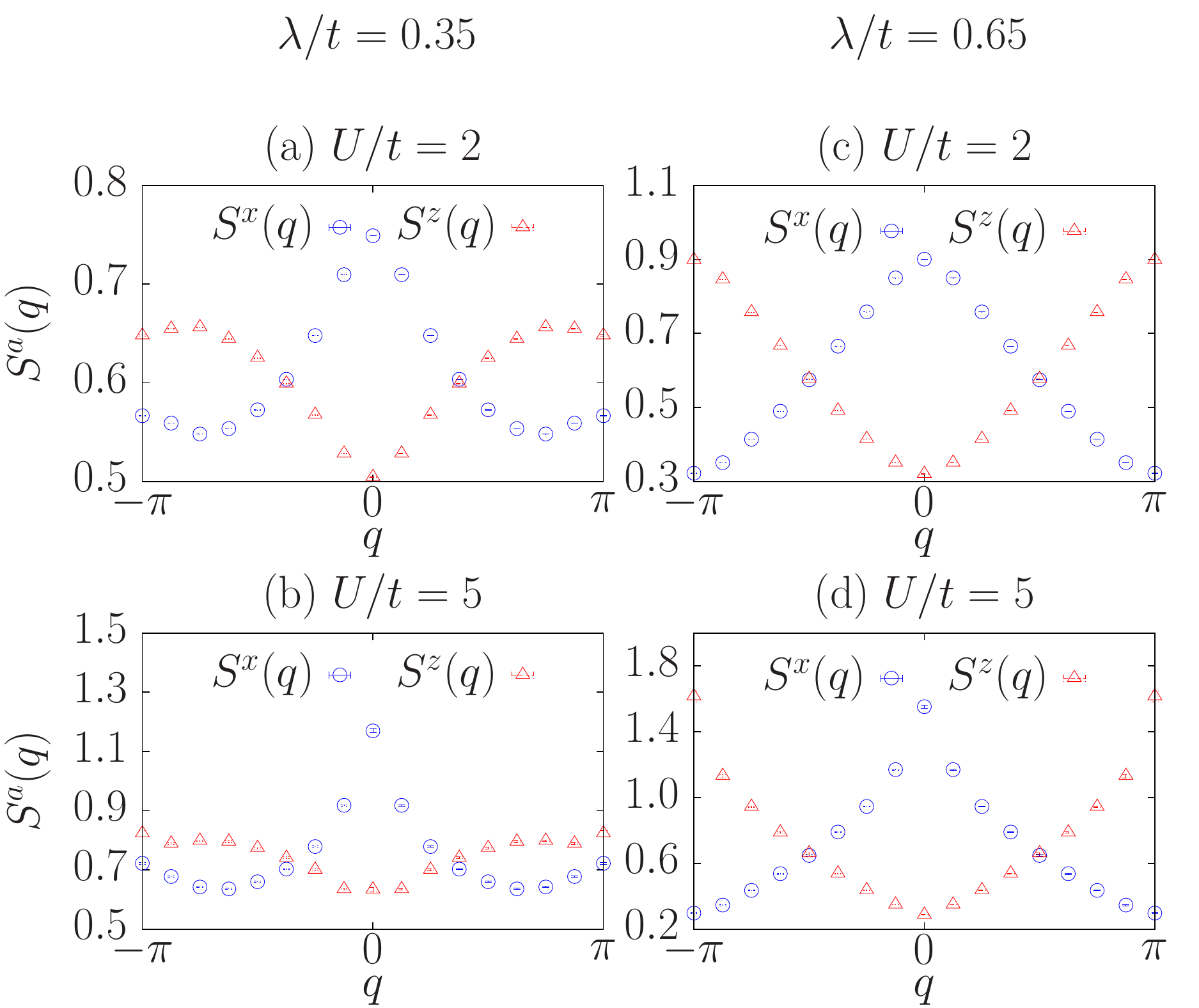}
  \caption{(Color online) Spin structure factors $S^{x}(q)$ and $S^{z}(q)$ [Eq.~(\ref{eq:sq})] from CT-INT simulations for $\lambda/t=0.35$ [(a),(b)] and 
    $\lambda/t=0.65$ [(c),(d)]. Here, $\lambda_\text{R}/t=0.3$. \label{fig_ctqmc6}}
\end{figure}

The spin structure factors ($a=x,z$)
\begin{equation}\label{eq:sq}
S^{a}(q)=\frac{1}{\sqrt{N}}\sum\limits_{x} e^{-iqx}\langle \hat{S}^{a}(x) \hat{S}^{a}(0)\rangle
\end{equation}
are shown in Fig.~\ref{fig_ctqmc6}. For $\lambda/t=0.65$ and $U/t=2$,
$S^{x}(q)$ has cusps at $q=0$ and $q=\pi$ [Fig.~\ref{fig_ctqmc6}(c)], and
varies almost linearly in between. With increasing $U/t$
[$U/t=5$ in Fig.~\ref{fig_ctqmc6}(d)], correlations with $q=0$ become much
stronger. Whereas $q=0$ spin correlations dominate the $x$ component of spin,
the structure factor $S^z(q)$ in Fig.~\ref{fig_ctqmc6}(d) indicates equally
strong correlations with $q=\pi$ for the $z$ component. The resulting spin
order resembles that of a canted antiferromagnet. Qualitatively similar
results, although with a less pronounced increase of spin correlations
between $U/t=2$ and $U/t=5$, are also observed for $\lambda/t=0.35$, as shown
in Figs.~\ref{fig_ctqmc6}(a),(b).

Despite a small but nonzero Rashba coupling, the results in
Figs.~\ref{fig_ctqmc6}(c) and (d) reveal the  symmetry relation
$S^{z}(q)=S^{x}(q+\pi)$ which roots in the chiral $SU(2)$ symmetry of the
corresponding low-energy Hamiltonian (see Sec.~\ref{subsec_symmetry}). Our
quantum Monte Carlo results show that this symmetry survives even in the
presence of strong correlations. The results in Fig.~\ref{fig_ctqmc6} are
almost identical to the case with $\lambda_\text{R}=0$ (not shown),
suggesting that the Rashba term breaks the chiral symmetry only weakly.
On the other hand, the symmetry is clearly absent for $\lambda/t=0.35$
[Figs.~\ref{fig_ctqmc6}(a),(b)].
\subsection{Effective spin model for $\lambda/t=\lambda_{s}$}
For strong interactions $U/t$, there exist no low-energy charge fluctuations
at the edge, allowing for a description in terms of a spin model.
We consider the case of (nearly) equal velocities, $\lambda/t=0.65$,
and make an ansatz in the form of a Heisenberg model with nearest-neighbor interactions, 
\begin{eqnarray}
\label{eqn_xxz}
\mathcal{H}_{\text{spin}}&=& \sum\limits_{i} \left(J_{x }S_{i}^{x}S_{i+1}^{x}+J_{y} S_{i}^{y}S_{i+1}^{y} + J_{z}S_{i}^{z}S_{i+1}^{z}\right)\nonumber\\
&=& J\sum\limits_{i} \left(S_{i}^{x}S_{i+1}^{x}+S_{i}^{y}S_{i+1}^{y}-S_{i}^{z}S_{i+1}^{z}\right)\;.
\end{eqnarray}
In the second line, the coupling constants $J_{a}$ have been fixed by
imposing the invariance under the rotations given in Eq.~(\ref{eqn_rot}),
$[H_{\text{spin}}^{s},U_{a}]=0$, and using the relations
$U_{a}^{\dagger}\hat{S}^{b}(x)U_{a}=M_{ab}$ [cf. Eq.~(\ref{eqn_rot2})]. Hamiltonian~(\ref{eqn_xxz}) corresponds to the XXZ
Heisenberg model, tuned to the ferromagnetic isotropic point that separates
the Ising phase from the  Luttinger liquid phase via a first order transition.
In both cases, one expects strong spin correlations, as observed in  Fig.~\ref{fig_ctqmc6}(d) \cite{Luther75}.
\section{Conclusions} \label{sec_sum}
In this paper, we introduced the $\pi$KM model, corresponding to the
Kane-Mele model on a honeycomb lattice with a magnetic flux of $\pm\pi$
through each hexagon. The flux insertion doubles the size of the unit cell,
and leads to a four-band model for each spin sector. For one spin direction,
the band structure has four Dirac points which acquire a gap for nonzero
spin-orbit coupling $\lambda$. At half filling, the spinless model has a Chern
insulating ground state with Chern number 2 or $-2$, depending on the
spin-orbit coupling. The transition between these states occurs via a phase
transition at $\lambda/t=1/2$, and the band structure features a quadratic
crossing at the critical point. The spinful $\pi$KM model is trivial in the $Z_2$
classification, with an even number of Kramers doublets. If translation
symmetry at the edge is unbroken, the helical edge states are stable at the
single-particle level even in the presence of a Rashba coupling that breaks
the $U(1)$ spin symmetry.
The $U(1)$ spin symmetric low-energy model of the edge states has a chiral symmetry 
when the edge state velocities have equal magnitude and either the same or
opposite sign. This chiral symmetry is shown to survive even in the presence
of interactions.

Regarding the effect of electronic correlations in the bulk, the combination of mean-field
calculations and quantum Monte Carlo simulations suggest the existence of a
quantum phase transition to a state with long-range, antiferromagnetic order,
similar to the Kane-Mele-Hubbard model. The critical value of the interaction
depends on the spin-orbit coupling. At $\lambda/t=1/2$, where the quadratic
band crossing occurs, a weak-coupling Stoner instability exists.

We studied the correlation effects on the edge states in the paramagnetic
bulk phase. At half filling, the bosonization analysis predicts the
opening of a gap in the edge states as a result of umklapp scattering for any
nonzero interaction. For strong coupling, we were able to confirm this
prediction using quantum Monte Carlo simulations. 
Umklapp processes are only effective at commensurate filling  and therefore
can be eliminated by doping away from half filling. 
In this case, we expect the interacting model to have stable edge modes, provided translation symmetry is not broken.
 At large $U/t$, the emergent chiral symmetry can be used to derive an effective
spin model of the XXZ Heisenberg type.

Our model may be regarded as a two-dimensional counterpart of TCIs. Whereas
the gapless edge states of the latter are protected by crystal symmetries of
the two-dimensional surface, the edge states in the $\pi$KM model are
protected (at the single-particle level, or away from half filling) by
translation symmetry.
TCIs have an even number of surface Dirac cones which are related by a crystal symmetry.
The cones can be displaced in momentum space without breaking time-reversal
symmetry by applying inhomogeneous strain \cite{Tang14}. This is in contrast to 
topological insulators with an odd number of Dirac points where at least one
Kramers doublet is pinned at a time-reversal invariant momentum. 
In TCIs, umklapp scattering processes can be avoided either by doping away from half filling 
or by moving the Dirac points.
In our model, the edge modes have in general unequal velocities and
cannot be mapped onto each other by symmetry. The Dirac points are
pinned at the time-reversal invariant momenta, and subject to umklapp
scattering at half filling.

Finally, the $\pi$KM model may be experimentally realized in ultracold atomic
gases by using optical flux lattices to create periodic magnetic flux densities
\cite{Goldman10,Cooper11,Aidelsburger11,Baur14,Celi14}.
{\begin{acknowledgments}%
We thank F.~Crepin and B.~Trauzettel for helpful discussions. We acknowledge
computing time granted by the J\"ulich Supercomputing Centre (JUROPA), and the Leibniz Supercomputing
Centre (SuperMUC). This work was supported by the DFG grants Nos. AS120/10-1
and Ho 4489/2-1 (FOR1807).
\end{acknowledgments}}
\bibliographystyle{apsrev4-1}
\bibliography{bercx2012}
\end{document}